\documentclass[sigconf, nonacm]{acmart}

\usepackage{amsmath}
\usepackage{amsthm}
\usepackage{multicol}
\usepackage{comment}
\usepackage{array}
\usepackage{eqparbox}
\usepackage{color}
\usepackage{subcaption}
\usepackage{caption} 
\usepackage{xspace}
\usepackage{slashbox}
\usepackage{float}
\usepackage{paralist}
\usepackage[linesnumbered,ruled,vlined]{algorithm2e}
\newcommand{\vect}[1]{\mathbf{#1}}
\newcommand{\size}[1]{|#1|}

\newcommand{\intersect}{\cap}
\newcommand{\setdiff}{\backslash}

\newcommand{\conj}{\wedge}

\newcommand{\set}[1]{\{\, #1 \,\}}

\newcommand{\revise}[1]{#1}

\newcommand{\mycomment}[1]{}
\newcommand{\mysubsection}[1]{\vspace{1ex}\noindent\textbf{#1.}\xspace}

\newtheorem{problem}{Problem}

\newcommand{\topk}{top-$k$\xspace}

\newcommand{\pexeso}{\textsf{PEXESO}\xspace}
\newcommand{\deepjoin}{\textsf{DeepJoin}\xspace}
\newcommand{\deepjoindbert}{$\textsf{DeepJoin}_\textsf{DistilBERT}$\xspace}
\newcommand{\deepjoinmpnet}{$\textsf{DeepJoin}_\textsf{MPNet}$\xspace}
\newcommand{\josie}{\textsf{JOSIE}\xspace}
\newcommand{\lshensemble}{\textsf{LSH Ensemble}\xspace}
\newcommand{\fasttext}{\textsf{fastText}\xspace}
\newcommand{\bert}{\textsf{BERT}\xspace}
\newcommand{\tabert}{\textsf{TaBERT}\xspace}
\newcommand{\turl}{\textsf{TURL}\xspace}
\newcommand{\mpnet}{\textsf{MPNet}\xspace}
\newcommand{\mlp}{\textsf{MLP}\xspace}

\SetFuncSty{text}
\SetArgSty{text}
\SetKw{OR}{or}
\SetKw{AND}{and}
\SetKw{NOT}{not}
\SetKw{TRUE}{true}
\SetKw{FALSE}{false}
\SetKw{NULL}{nil}
\SetKw{Break}{break}
\SetKw{Continue}{continue}
\SetKwInOut{Input}{Input}
\SetKwInOut{Output}{Output}

\input{savespace}

\begin{document}
\title{\deepjoin: Joinable Table Discovery with Pre-trained Language Models}

%%
%% The "author" command and its associated commands are used to define the authors and their affiliations.
% \newcommand{\nec}{$^{1}$}
% \newcommand{\osanag}{$^{2,3}$}
% \newcommand{\osaka}{$^{2}$}
% \newcommand{\nagoya}{$^{3}$}

% \author{\nec Yuyang Dong, \osanag Chuan Xiao, \nec Takuma Nozawa, \nec Masafumi Enomoto, \nec Masafumi Oyamada}

% \affiliation{\nec NEC Corporation, \osaka Osaka University, \nagoya Nagoya University}

% \email{{dongyuyang, nozawa-takuma, masafumi-enomoto, oyamada}@nec.com, chuanx@ist.osaka-u.ac.jp}
 
\author{Yuyang Dong}
\affiliation{%
  \institution{NEC Corporation}
  %\country{Japan}
}
\email{dongyuyang@nec.com}

\author{Chuan Xiao}
\affiliation{%
  \institution{Osaka University}
  \institution{Nagoya University}
  %\country{Japan}
}
\email{chuanx@ist.osaka-u.ac.jp}

\author{Takuma Nozawa}
\affiliation{%
  \institution{NEC Corporation}
  %\country{Japan}
}
\email{nozawa-takuma@nec.com}

\author{Masafumi Enomoto}
\affiliation{%
  \institution{NEC Corporation}
  %\country{Japan}
}
\email{masafumi-enomoto@nec.com}

\author{Masafumi Oyamada}
\affiliation{%
  \institution{NEC Corporation}
  %\country{Japan}
}
\email{oyamada@nec.com}

\begin{abstract}
%   Given a table repository and a query table with a specified join column, joinable 
%   table discovery finds the target tables that can be joined with the query. 
  Due to the usefulness in data enrichment for data analysis tasks, joinable table 
  discovery has become an important operation in data lake management. Existing 
  approaches target equi-joins, the most common way of combining tables for creating 
  a unified view, or semantic joins, which tolerate misspellings and different 
  formats to deliver more join results. They are either exact solutions whose 
  running time is linear in the sizes of query column and target table repository, or 
  approximate solutions lacking precision. In this paper, we propose \deepjoin, a 
  deep learning model for accurate and efficient joinable table discovery. Our 
  solution is an embedding-based retrieval, which employs a pre-trained language 
  model (PLM) and is designed as one framework serving both equi- and semantic (with 
  a similarity condition on word embeddings) joins for textual attributes with fairly 
  small cardinalities. 
  We propose a set of contextualization options to transform column contents 
  to a text sequence. The PLM reads the sequence and is fine-tuned to embed columns to 
  vectors such that columns are expected to be joinable if they are close to each 
  other in the vector space. Since the output of the PLM is fixed in length, the 
  subsequent search procedure becomes independent of the column size. With a 
  state-of-the-art approximate nearest neighbor search algorithm, the search time is 
  sublinear in the repository size. To train the model, we devise the techniques 
  for preparing training data as well as data augmentation. The experiments on real 
  datasets demonstrate that by training on a small subset of a corpus, \deepjoin 
  generalizes to large datasets and its precision consistently outperforms other 
  approximate solutions'. \deepjoin is even more accurate than an exact solution to 
  semantic joins when evaluated with labels from experts. Moreover, when equipped with 
  a GPU, \deepjoin is up to two orders of magnitude faster than existing solutions. 
\end{abstract}

\maketitle

\section{Introduction}
\label{sec:intro}

Given a table repository and a query table with a specified join column, joinable 
table discovery finds the target tables that can be joined with the query. Due to 
the demonstrated usefulness in data enrichment~\cite{arda, table-enrichment}, 
joinable table discovery has become a key procedure in data lake management and 
serves various downstream applications, especially those involving data analysis. 

For joinable table discovery, early attempts mainly targeted 
equi-joins~\cite{lsh-ensemble, josie}, which are the most common way of combining 
tables for creating a unified view~\cite{data-integration-book} and can be easily 
implemented using SQL. To deliver more joins results for heterogeneous data, recent 
approaches~\cite{pexeso, table-enrichment} studied semantic joins, which join on  
cells with similar meanings via word embedding, so as to handle data with 
misspellings and discrepancy in formats/terminologies (e.g., ``American Indian \& Alaska 
Native'' v.s. ``Mainland Indigenous''). There are two major limitations in these 
solutions. First, they only apply to a single join type. Second, most of them are 
exact algorithms with a worst-case time complexity linear in the product of query 
column size and table repository size, and thus their scalability is dubious. 
Despite the existence of an approximate algorithm for equi-joins~\cite{lsh-ensemble}, 
it is based on MinHash sketches~\cite{minhash} and has to convert the joinability 
condition to a Jaccard similarity condition, which is non-equivalent and introduces 
many false positives. Moreover, it is sometimes even slower than an exact 
algorithm~\cite{josie}. 

Seeing the limitations of existing solutions, we propose \deepjoin, a deep learning 
model designed in a \emph{two-birds-with-one-stone} fashion such that both equi- and 
semantic joins can be served with one framework. In particular, \deepjoin targets 
textual attributes with fairly small cardinalities that can fit with language models. 
To cope with semantic joins, it works on a similarity condition of word embeddings and 
finds similar textual columns as exactly as possible. To resolve the efficiency 
issue, it finds joinable tables via an \textbf{embedding-based retrieval}. In 
particular, we employ an embedding model to transform columns to a vector space. By 
metric learning, columns with high joinability are close to each other in the vector 
space. Then, to find the \topk target columns ranked by joinability, we resort to a
state-of-the-art approximate nearest neighbor search (ANNS) algorithm~\cite{hnsw}, 
whose time complexity is \emph{sublinear} in the table repository size.

\deepjoin utilizes a \textbf{pre-trained language model} (PLM) for column embedding. 
PLMs, such as BERT~\cite{bert}, have gained popularity in various data management 
tasks that involve natural language processing. A salient property of modern PLMs is 
that they are transformer networks~\cite{transformer} featuring the attention 
mechanism, thus not only good at capturing the semantics of column contents for 
semantic joins, but also able to focus on the cells that are more probable to match in 
equi-joins, assuming that the query column has a similar distribution to those in 
the repository. As such, our model gains the capability of handling both join types, 
only needing the PLM to be fine-tuned on the data labeled for either equi- or semantic 
joins. In addition, PLMs produce a fixed-length vector, meaning that the following 
index lookup and search are \emph{independent} of the column size, hence along with the 
ANNS, addressing the scalability issue in existing solutions. Since PLMs take a text 
sequence as input, by prompt engineering, we propose a set of options that 
contextualizes a column to a text sequence. To train the model in a self-supervised 
manner, we devise a series of techniques to effectively generate positive and negative 
examples. Moreover, our training features a data augmentation technique, through which 
our model can learn that the joinability is insensitive to the order of cells in a 
column. 

We conduct experiments on two real datasets and evaluate \deepjoin equipped with two 
state-of-the-art PLMs. We show that by training on a small subset (30k columns) of 
the corpus, \deepjoin \emph{generalizes} well to large datasets (1M columns). In 
particular, \deepjoin outperforms alternative approximate solutions in all the 
settings, and reports an average precision of 72\% for equi-joins and 91\% for 
semantic joins and an average NDCG of 81\% for equi-joins and 75\% for semantic joins. 
To test the effectiveness of semantic joins, we also evaluate \deepjoin using data 
labeled by our database researchers, and the results show that \deepjoin is even better 
by a margin of 0.105 -- 0.165 F1 score than \pexeso~\cite{pexeso}, an exact solution 
we use to label \deepjoin's training data. An ablation study demonstrates the 
usefulness of the proposed contextualization and data augmentation techniques. For 
scalability test, we vary dataset size from 1M to 5M columns. Even if equipped with a 
CPU, \deepjoin exhibits superb scalability and is 7 -- 57 times faster than existing 
solutions. With the help of a GPU, \deepjoin outperforms them by up to two orders of 
magnitude. 

\mysubsection{Contributions}
% Our contributions are summarized as follows. 
\begin{inparaenum} [(1)]
  \item We propose \deepjoin, a framework for joinable search discovery in a data lake. 
  Our solution targets targets textual attributes with fairly small cardinalities, and 
  is able to detect both equi-joinable and semantically joinable tables. 
  \item We design the search in \deepjoin as an embedding-based retrieval  
  which employs a fine-tuned PLM for column embedding and ANNS for fast retrieval. The 
  search time complexity is sublinear in the table repository size, and except the 
  column embedding, the search time is independent of the column size. 
%   Through metric learning, the PLM is able to embed columns to vectors such that 
%   columns are expected to be joinable if they are close in the vector space. 
  \item We propose a prompt engineering method to transform column contents to a text 
  sequence fed to the PLM. 
  \item We devise techniques for training data preparation and data augmentation, as 
  well as a metric learning method to fine-tune the PLM in a self-supervised manner. 
  \item We conduct experiments to show that our model generalizes well to large datasets 
  and it is accurate and scalable in finding joinable tables. 
\end{inparaenum}

Furthermore, we would like to mention the following facts: 
\begin{inparaenum} [(1)]
  \item For the embedding-based retrieval in our model, the results of ANNS are directly 
  output as the results of joinable table discovery, whereas more advanced paradigms 
  exist, such as two-stage retrieval~\cite{youtube-recsys} which finds a set of 
  candidates by ANNS and ranks the candidates by a more sophisticated model. 
  \item \deepjoin is not limited to the two PLMs evaluated in our experiments, because 
  the PLM can be regarded as a plug-in in our framework. 
\end{inparaenum}
As such, we expect the performance of \deepjoin can be further improved by using more 
advanced retrieval paradigms or PLMs.

\section{Preliminaries}
\label{sec:prelim}

% In this section, we present a formal definition of the joinable 
% table discovery problem and briefly review related work on pre-trained 
% language models. 

% \begin{table} [t]
%   \small
%   \centering
%   \caption{Notation used in this paper.}  
%   \resizebox{\linewidth}{!}{%
%   \begin{tabular}{|l|l|} \hline
%     Symbol & Description \\ \hline \hline
%     $Q$ & query column \\ \hline
%     $X$ & target column in the repository \\ \hline    
%     $\mathcal{X}$ & collection of target columns (i.e., the repository) \\ \hline
%     %$\vect{q}, \vect{x}$ & embeddings of $Q$ and $X$, respectively \\ \hline
%     $jn(Q, X)$ & joinability from $Q$ to $X$ \\ \hline
%     %$d(\cdot, \cdot)$ & (Euclidean) distance between embeddings \\ \hline
%     $k$ & number of results \\ \hline
%     $c$ & number of candidate target columns \\ \hline
%   \end{tabular}
%   }
%   \label{tab:notation}
% \end{table}

\subsection{Problem Definition}
\label{sec:problem-definition}
%We present a formal definition of the joinable table discovery problem. 
%Table~\ref{tab:notation} summarizes the notation used in this paper. 

Given a data lake of tables, we extract all the columns in these tables, 
except those unlikely to appear in a join predicate (e.g., BLOBs), and 
create a repository of tables, denoted by $\mathcal{X}$. Given a query 
column $Q$, our task is to search $\mathcal{X}$ and find the columns 
joinable to $Q$. In this paper, we target equi-joins and semantic joins. 
Next, we define the joinability for these two types, respectively. 
% Equi-join is the most common way of combining tables for 
% create a unified view~\cite{data-integration-book} and can be easily 
% implemented using SQL, while semantic join is defined over word 
% embeddings and is able to handle data with misspellings and different 
% formats/terminologies (e.g., ``American Indian \& Alaska Native'' v.s. 
% ``Mainland Indigenous''), hence delivering more join results.  
% Moreover, transformation tools~\cite{auto-join,auto-transform} are available 
% to convert tables for subsequent equi-joins in case they differ in format. 

Given a query column $Q$ and a target column $X$ in $\mathcal{X}$, 
the joinability from $Q$ to $X$ is defined by the following equation. 
\begin{align}
    \label{eq:joinability}
    jn(Q, X) = \frac{\size{Q_M}}{\size{Q}}, 
\end{align}
where $\size{\cdot}$ measures the size (i.e., the number of cells) of a 
column, and $Q_M$ is composed of the cells in $Q$ that have at least one 
match in $X$. Here, the term ``match'' depends on the join type, i.e., an 
equi-join or a semantic join. We normalize the size of $Q_M$ by the size 
of $Q$ to return a value between 0 and 1. Moreover, the joinability is 
not always symmetric, depending on the definition of $Q_M$. 

For equi-joins, we model each column as a set of cells by removing 
duplicate cell values, and define the equi-joinability as follows.

\begin{definition} [Equi-Joinability]
  \label{def:equi-joinablity}
  The equi-joinability from the query column $Q$ to a target column $X$ in 
  $\mathcal{X}$ counts the intersection between $Q$ and $X$, normalized by 
  the size of $Q$; i.e., in Equation~\ref{eq:joinability}, 
  \begin{align}
    \label{eq:equi-joinability}
    Q_M = Q \intersect X. 
  \end{align}
\end{definition}

The equi-joinability defined above counts the the distinct number of cells in 
$Q$ that match those in $X$, and thus can be used to measure the 
equi-joinability~\cite{josie}. 
Our method can be also extended to the case when columns are modeled as 
multisets, so as to support one-to-many, many-to-one, and many-to-many joins. 
In this case, we may measure the joinability by the number of join results 
and normalize it by the product of $\size{Q}$ and $\size{X}$, instead of 
$\size{Q}$ in Equation~\ref{eq:joinability}. 

For semantic joins, we consider string columns and embed the value of each cell 
to a metric space $\mathcal{V}$ (e.g., word embedding by fastText~\cite{fasttext}). 
As such, each string column is transformed to a multiset of vectors. We abuse the 
notation of a column to denote its multiset of vectors. Then, the notion of vector 
matching is defined as follows. 

\begin{definition} [Vector Matching]
  \label{def:vector-match}
  Given two vectors $v_1$ and $v_2$ in $\mathcal{V}$, a distance function $d$, 
  and a threshold $\tau$, $v_1$ matches $v_2$ if and only if 
  the distance between $v_1$ and $v_2$ does not exceed $\tau$. We use notation 
  $M_\tau^d(v_1, v_2)$ to denote if $v_1$ matches $v_2$; i.e., 
  $M_\tau^d(v_1, v_2) = 1$, iff. $d(v_1, v_2) \leq \tau$, or 0, otherwise. 
\end{definition}

Given a query column $Q$ and a target column $X$, the semantic-joinability is 
defined using the number of matching vectors~\footnote{Besides this definition,  
semantic joins are also investigated in \cite{sema-join}, yet it studies the 
problem of performing joins rather than searching for joinable columns.}. 

\begin{definition} [Semantic-Joinability]
  \label{def:semantic-joinability}
  The semantic-joinability from $Q$ to $X$ counts the number of vectors in $Q$ 
  having at least one matching vector in $X$, normalized by the size of $Q$; 
  i.e., in Equation~\ref{eq:joinability}, 
  \begin{align}
    \label{eq:semantic-joinability}
    %& jn_{\text{semantic}}(Q, X) = \frac{|Q_M|}{|Q|}, \\
    Q_M = \set{q \mid q \in Q \conj \exists x \in X \text{\, s.t.\, } M_\tau^d(q, x) = 1}.
  \end{align}
\end{definition}

% Besides equi-joinability, there are also approaches to enriching the join 
% result by identifying semantic-joinable results~\cite{sema-join,pexeso}, yet 
% these approaches either perform joins rather than search for joinable 
% columns~\cite{sema-join} or require users to operate on the word embeddings with 
% a user-defined function to specify the join predicate, thereby less convenient 
% than using equi-joinability.

An advantage of the above definitions is that for both equi- and 
semantic-joinability, the training data can be labeled by an exact algorithm 
(e.g., \josie~\cite{josie} and \pexeso~\cite{pexeso}) %(see Section~\ref{sec:train}) 
rather than experts, so that our model can be trained in a self-supervised 
manner. Following the above definitions, we model the problem of joinable 
table discovery as the following \topk search problem, where joinability 
$jn$ is defined using either Definition~\ref{def:equi-joinablity} or 
\ref{def:semantic-joinability}. 

\begin{problem} [Joinable Table Discovery]
  \label{pb:joinable-table-discovery}
  Given a query column $Q$ and a repository of target columns $\mathcal{X}$, the 
  joinable table discovery problem is to find the \topk columns in $\mathcal{X}$ 
  with the highest joinability from $Q$. Formally, 
  we find a subset 
  $\mathcal{R} \subseteq \mathcal{X}$, $\size{\mathcal{R}} = k$, and 
  $\min\set{jn(Q, X) \mid X \in \mathcal{R})} \geq jn(Q, Y), 
  \forall Y \in \mathcal{X} \setdiff \mathcal{R}$.
\end{problem}

\revise{
In this paper, we focus on dealing with textual columns. For numerical columns, 
a typical solution is utilizing the statistical feature vector in 
Sherlock~\cite{sherlock}, which converts a numerical column to a vector, and then 
we can invoke a vector search to look for joinable columns having similar statistics 
to the query column. 
}

\subsection{State-of-the-Art}
\label{sec:state-of-the-art}
\josie~\cite{josie}, a state-of-the-art solution to the equi-joinable table 
discovery problem, regards the problem as a \topk set similarity search with 
overlap $\size{Q \intersect X}$ as similarity function, and builds its 
search algorithm upon prefix filter~\cite{prefix-filter} and positional 
filter~\cite{positional-filter}, which have been extensively used for solving 
set similarity queries. 
% \josie works in the following steps: 
% \begin{inparaenum} [(1)]
%   \item Create an inverted index $\mathcal{X}$, which regards each cell value 
%   as a token and maps each token to a postings list of columns having the token. 
%   \item Sort the tokens in each column by a global order, and find a set of 
%   candidate columns by retriving the postings lists for the first few tokens of 
%   $Q$ (called the prefix of $Q$). 
%   \item Compute the overlap for the candidates and update the \topk results. 
% \end{inparaenum}
\josie creates an inverted index over $\mathcal{X}$, which regards each cell 
value as a token and maps each token to a postings list of columns having the 
token. Then, it finds a set of candidate columns by retrieving the postings lists 
for a subset of the tokens in $Q$ (called prefix). Candidates are verified for 
joinability and the \topk is updated. While index access and candidate verification 
are processed in an alternate manner, \josie features the techniques to 
determine the their order, so as to optimize for long columns and large token 
universes, which are often observed in data lakes. 
% While steps (2) and (3) are processed in an alternate manner, a prefix length 
% can be inferred from the running $k$-th result's overlap, so as to guarantee an 
% exact solution to the problem. Moreover, \josie features the techniques to 
% determine the order of processing (2) and (3) to optimize for large columns and 
% large token universes, which are often observed in data lakes. 

\pexeso~\cite{pexeso} is an exact solution to semantic-joinable table discovery. 
It employs pivot-based filtering~\cite{pivot-filtering}, which selects a set of 
vectors as pivots and pre-computes distances to these pivots for all the vecotrs 
in the columns of $\mathcal{X}$. Then, given the vectors of the query $Q$, 
non-matching vectors can be pruned by the triangle inequality. A hierarchical 
grid index is built to filter non-joinable columns when counting the number of 
matching vectors. 

As for the weakness, \josie inherits the limitation of prefix filter, whose 
performance highly depends on the data distribution and yields a worst-case 
time complexity of $O(\size{\mathcal{X}} \cdot (\size{Q} + \overline{\size{X}}))$, 
where $\overline{\size{X}}$ stands for the average size of the columns in 
$\mathcal{X}$. For \pexeso, despite a claimed sublinear search time complexity of 
$O(\log\size{\mathcal{X}_V} \cdot \log\size{Q})$, where $\mathcal{X}_V$ denotes 
the multiset of all the vectors in the repository, it relies on a 
user-specific threshold for the count of matching vectors. This does not apply to 
the \topk case, and the algorithm is downgraded to be linear in 
$\size{\mathcal{X}_V} \cdot \size{Q}$, because at the early stage of search, 
due to the low count of matching vectors in the temporary \topk results, the pruning 
power of the grid index is next to none. In general, for search time, both \josie 
and \pexeso are linear in the product of column size and repository size, which 
compromises their scalability to long columns and large datasets. On the other hand, 
it is unnecessary to always find an exact answer, because data scientists are 
usually concerned with only part of the \topk results and will choose a subset of 
them for the join. For this reason, we will design our solution with the following 
two goals: 
\begin{inparaenum} [(A)]
  \item it returns an approximate answer with \emph{sublinear} time in 
  $\size{\mathcal{X}}$, and 
  \item by encoding the query column to a fixed length, it is 
  \emph{independent} of $\size{Q}$ and $\overline{\size{X}}$ during index lookup and 
  search. 
\end{inparaenum}

% Note that ``approximate'' does not mean provable guanratee. 
% We use the term in line with the studies on $k$-nearest neighbor 
% search~\cite{knn-tutorial} that seek approximate answers. 

Apart from exact solutions, \lshensemble~\cite{lsh-ensemble} is an approximate 
solution to equi-joinability. It partitions the repository and computes MinHash 
sketches~\cite{minhash} with parameters tuned for each partition. Unlike \josie, 
it targets the thresholded problem (i.e., 
$\frac{\size{Q \intersect X}}{\size{Q}} \geq t$), which requires a user-specified 
threshold $t$ and thus is less flexible than computing \topk. Although adaptation 
for the \topk problem is available, it suffers from low precision due to the many 
false positives introduced by transforming overlap similarity to Jaccard 
similarity for the use of MinHash, and it sometimes runs even slower than 
\josie~\cite{josie}.

\section{The \deepjoin Model}
\label{sec:model}

\begin{figure}[t]
  \centering
  \includegraphics[width=\linewidth]{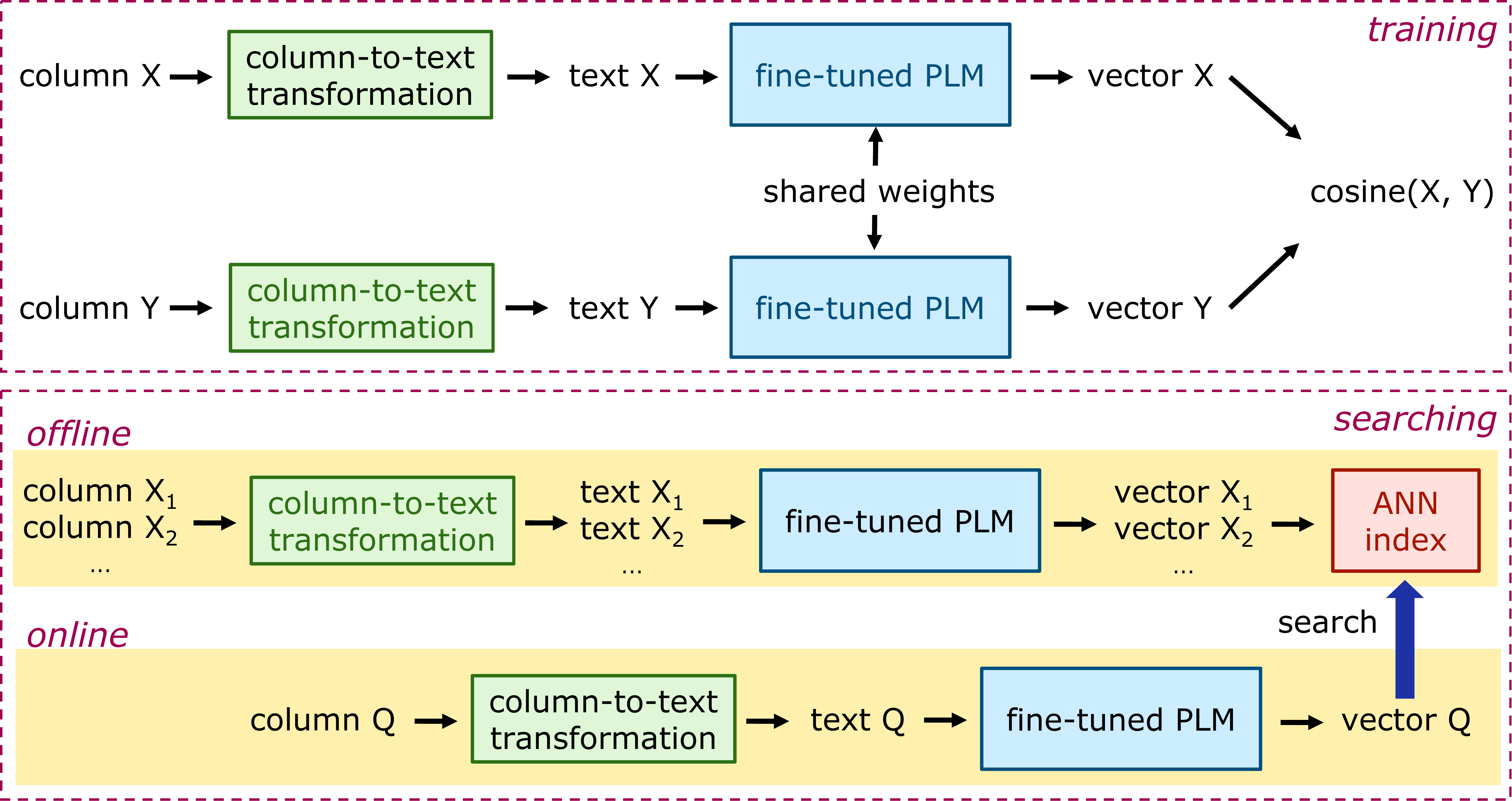}
  \caption{Overview of the \deepjoin model.}
  \label{fig:overview}
\end{figure}

% \begin{figure*}[t]
%   \centering
%   \includegraphics[width=0.8\textwidth]{figures/overview-crop.pdf}
%   \caption{Overview of joinable table search with \deepjoin, offline indexing and online searching.}
%   \label{fig:overview}
% \end{figure*}

\begin{table} [t]
  \small
  \centering
  \caption{Column-to-text transformation options.}
  \resizebox{\linewidth}{!}{%
  \begin{tabular}{|l|l|} \hline
    Name & Pattern \\ \hline \hline
    \texttt{col} & \$\texttt{cell\_1}\$, \$\texttt{cell\_2}\$, $\ldots$, \$\texttt{cell\_n}\$ \\ \hline
    \texttt{colname-col} & \$\texttt{column\_name}\$ : \$\texttt{col}\$. \\ \hline    
    \texttt{colname-col-context} & \$\texttt{colname-col}\$ : \$\texttt{col}\$. \$\texttt{table\_context}\$ \\ \hline
    \texttt{colname-stat-col} & \$\texttt{column\_name}\$ contains \$\texttt{n}\$ values \\
    & (\$\texttt{max\_len}\$, \$\texttt{min\_len}\$, \$\texttt{avg\_len}\$): \$\texttt{col}\$  
    \\ \hline
    \texttt{title-colname-col} & \$\texttt{table\_title}\$. \$\texttt{colname-col}\$ \\ \hline
    \texttt{title-colname-col-context} & \$\texttt{title-colname-col}\$. \$\texttt{table\_context}\$ \\ \hline
    \texttt{title-colname-stat-col} & \$\texttt{table\_title}\$. \$\texttt{colname-stat-col}\$. \\ \hline
  \end{tabular}
  }
  \label{tab:column-to-text}
\end{table}

% Figure~\ref{fig:overview} illustrates the overview of the \deepjoin model. 
% The design of the model is inspired by modern recommender systems that adopt 
% a two-stage paradigm~\cite{youtube-recsys}, which has been used in various 
% online services such as video sharing~\cite{youtube-recsys}, social 
% networking~\cite{facebook-recsys}, and online shopping~\cite{alibaba-recsys}. 
% The paradigm consists of a candidate generation (a.k.a. retrieval) stage and 
% a ranking stage. The candidate generation stage retrieves a small subset of 
% candidate columns from repository. The ranking stage ranks the candidates in 
% the order of descending joinability to the query column. Next, we present the 
% candidate generation in \deepjoin. 

%\subsection{Candidate Generation}

\begin{figure*}[t]
  \centering
  \includegraphics[width=\textwidth]{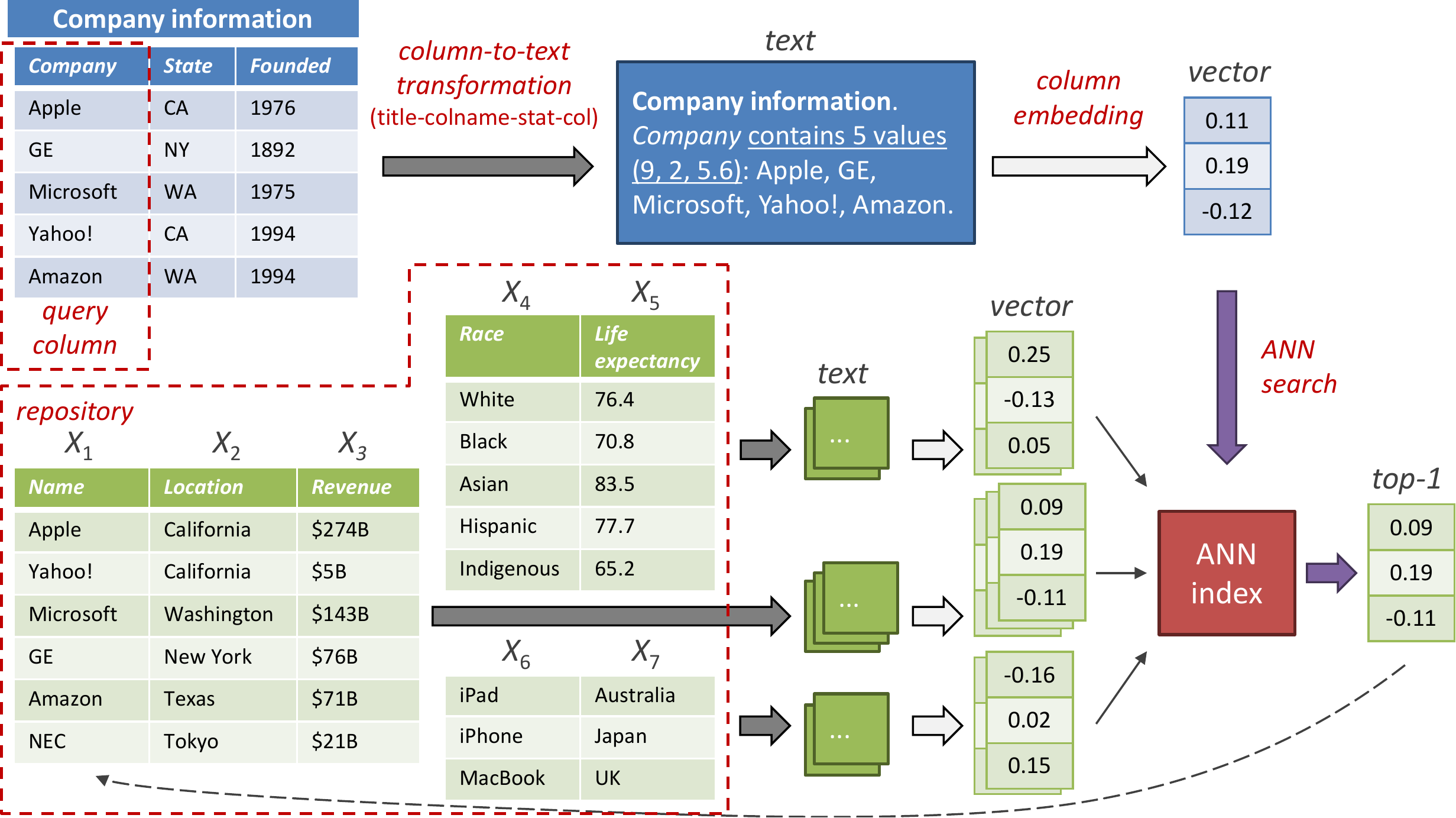}
  \caption{\revise{A running example of \deepjoin over a repository of 7 columns, $k = 1$.}}
  \label{fig:running-example}
\end{figure*}

Figure~\ref{fig:overview} illustrates the overview of our \deepjoin model. 
In \deepjoin, the joinable table discovery is essentially an embedding-based 
retrieval, which has recently been adopted in various retrieval tasks, 
such as long text retrieval~\cite{dpr} and search in social 
networks~\cite{facebook-recsys}. An embedding-based retrieval usually employs 
metric learning or meta embedding search to learn embeddings for target data 
such that the semantics can be measured in the embedding space, especially 
when target data are highly sparse or the semantics is hard to define by an 
analytical model. Another key benefit is, by embedding original data objects 
(i.e., columns) to a fixed-length vector, the subsequent procedure can be 
independent of the size of the data object, thereby achieving goal (B) stated 
in Section~\ref{sec:state-of-the-art}.

Since semantic joins are also in our scope, an immediate idea is employing a 
PLM to capture the semantics. 
% PLMs, such as BERT~\cite{bert}, have gained popularity in various data management tasks. 
By unsupervised training for 
language prediction tasks on a large text corpora such as Wikipedia pages, PLMs 
are able to embed texts to a vector space such that texts are expected 
to be similar in meaning if they are closer in the vector space. A key advantage 
of PLMs is that they are deep models that can be tailored to 
downstream tasks (e.g., data preparation~\cite{rpt}, entity 
matching~\cite{ditto}, and column annotation~\cite{doduo}) by fine-tuning with 
task-specific training data. Moreover, state-of-the-art PLMs  
% are mainly transformer networks~\cite{transformer}, which are deep neural networks that 
utilize the attention mechanism to focus on informative words than stop words. 
Besides handling the semantics, the attention mechanism can be also useful for 
identifying equi-joinable columns, because it can focus on the cells that are 
more probable to yield a match for equi-joins, assuming that the query column has 
a similar cell distribution to those in the repository. As such, we are 
able to use one model framework to cope with both equi-joins and semantic joins. 
The only difference is that the model is trained with data labeled for the 
target join type. In \deepjoin, we use a fine-tuned PLM to embed columns to a 
vector space such that columns with high joinability are close to each other in 
the vector space. Since PLMs take as input raw text, we transform (i.e., 
contextualize) the contents in each column to a text sequence, and then feed the 
sequence to the fine-tuned PLM to produce a column embedding. 

\subsection{Column-to-Text Transformation}
\label{sec:column-to-text}
The column-to-text transformation belongs to prompt engineering, which works 
by including the description of the task in the input to train a model 
and has shown effectiveness in natural language processing tasks such as 
question answering~\cite{cot}. In \deepjoin, we take advantage of metadata and 
consider seven options, shown in Table~\ref{tab:column-to-text}, where 
variables are quoted in dollar signs. In particular, \texttt{n} denotes the 
number of distinct cell values of the column; \texttt{cell\_i} denotes the 
value of the $i$-th cell of the column, with duplicate values removed; 
%\texttt{column\_name} denotes the column name. 
\texttt{stat} denotes the statistics of the column, including the maximum, 
minimum, and average numbers of words in a cell; and 
\texttt{table\_context} denotes the accompanied context of the table (e.g., a 
brief description). 
%\texttt{table\_title} denotes the title of the table. 
Some patterns are also used for creating other patterns. 
For example, \texttt{col} stands for the concatenation of all the cell values, 
with a comma as delimiter, and \texttt{colname-col} stands for the column name 
followed by a colon and the content of \texttt{col}. 

% In the above options, \texttt{column\_name}, \texttt{table\_context}, and 
% \texttt{table\_title} are metadata that might help improve the performance. Such 
% information is not always available, especially when csv files are loaded into 
% data lakes. \texttt{stat} provides statistical information of the column.

\revise{
\begin{example}
  \label{ex:column-to-text}
  Consider the query column in Figure~\ref{fig:running-example}. 
  Suppose \texttt{title-colname-stat-col} is used for column-to-text 
  transformation. The table title is ``\textsf{Company information}''. The 
  column name is ``\textsf{Company}''. There are 5 cell values: ``\textsf{Apple}'', 
  ``\textsf{GE}'', ``\textsf{Microsoft}'', ``\textsf{Yahoo!}'', and 
  ``\textsf{Amazon}'', with a maximum of 9 characters, a minimum of 2 characters, 
  and an average of 5.6 characters. Therefore, according to the patterns shown in 
  Table~\ref{tab:column-to-text}, the column is transformed to a text sequence 
  ``\textsf{Company information. Company contains 5 values (9, 2, 5.6): Apple, GE, 
  Microsoft, Yahoo!, Amazon.}'', as shown in Figure~\ref{fig:running-example}. 
  For the repository, we can transform the 7 target columns to 7 text 
  sequences using the same technique. 
\end{example}
}

\subsection{Column Embedding} 
\label{sec:embed-column}
In \deepjoin, we fine-tune two state-of-the-art PLMs: 
DistilBERT~\cite{distilbert}, a faster and lighter variant of BERT~\cite{bert}, 
and MPNet~\cite{mpnet}, which leverages the dependency among predicted tokens 
through permuted language modeling and takes auxiliary position information as 
input to make the model see a full sentence, thereby reducing position 
discrepancy. We use sentence-transformers~\cite{sbert} to output a sentence 
embedding for a sequence of input text. 
% Since \deepjoin is independent of the choice of PLM, 
% we expect its performance can be improved by more advanced PLMs. 
It is also noteworthy to mention that for semantic joins, unlike \pexeso, 
we do not need to generate embeddings in the vector space $\mathcal{V}$ 
(Definition~\ref{def:vector-match}). Instead, the semantics is captured by 
the fine-tuned PLM. 

\revise{Since PLMs have an input length limit \textsf{max\_seq\_length} 
(e.g., 512 tokens for BERT), in the case of a tall input column, we choose 
a frequency-based approach, e.g., taking a sample of the most frequent 
cell values from the column, whose number of tokens is just within 
\textsf{max\_seq\_length}. Then, we use the sample for column-to-text 
transformation. 
The reason is that they are more likely to 
yield join results. Here, the frequency of a cell value is defined as 
document frequency, i.e., the number of target columns in the repository 
that have this cell value. If columns are modeled as multisets, we may 
resort to other frequency-based approaches, such as TF-IDF and 
BM25~\cite{DBLP:books/daglib/0021593}.} 

\subsection{Indexing and Searching}
\label{sec:index-and-search}
In order to scale to large table repositories, we resort to approximate nearest 
neighbor search (ANNS). The embeddings of the columns in $\mathcal{X}$ are 
indexed offline. For online search, we find the $k$ nearest neighbors (kNN) 
of the query column embedding under Euclidean distance as search results. 
% , where $c \geq k$ denotes the number of candidates. 
In particular, we use hierarchical navigable small world (HNSW) \cite{hnsw}, 
which is among the most prevalent approaches to ANNS~\cite{knn-tutorial}. Since 
the search time complexity of HNSW is sublinear in the number of indexed 
objects~\cite{hnsw}, the search can be done with a time complexity sublinear in 
the number of target columns, thereby achieving goal (A) stated 
in stated in Section~\ref{sec:state-of-the-art}. Moreover, for billion-scale 
datasets, we may use an inverted index over product quantization (IVFPQ) 
\cite{ivfpq}, and construct HNSW over the coarse quantizer of IVFPQ. Such 
choice has become the common practice of billion-scale kNN search (e.g., using 
the Faiss library~\cite{faiss}). 

\revise{
\begin{example}
  \label{ex:index-and-search}
  Following Example~\ref{ex:column-to-text}, the text sequences for query and 
  target columns are embedded to vectors by a fine-tuned PLM, as shown in 
  Figure~\ref{fig:running-example}. The vectors for the 7 target columns are 
  indexed in the approximate nearest neighbor (ANN) index. Given the vector of 
  the query column, we search for its kNN ($k = 1$ in this example) by looking 
  up the ANN index. Then the result is output as the joinable column, which 
  refers to a target column in the repository. 
\end{example}
}

% \subsection{Ranking}
% In the ranking stage, we evaluate each candidate by computing the joinability 
% to $Q$ using Equation~\ref{eq:joinability}. 

\subsection{Complexity Analysis}
\label{sec:time-complexity}
The search consists of two parts: query encoding and ANNS. In query 
encoding, we transform the column to text, with a time complexity of 
$O(\size{Q})$, and then we feed the text to the PLM, with a time 
complexity of $O(\size{M} \cdot \size{Q})$, where $\size{M}$ denotes 
the model size. In ANNS, thanks to the use of HNSW, the time 
complexity~\footnote{We follow the complexity analysis in \cite{hnsw}.} 
is $O(vl\log\size{\mathcal{X}}$, where $v$ is the maximum out-degree 
(controlled by a parameter for index construction) in HNSW's index and 
$l$ is the dimensionality of column embedding. Compared to \josie and 
\pexeso, which are linear in 
$\size{\mathcal{X}} \cdot (\size{Q} + \overline{\size{X}})$, we reduce 
the time complexity to sublinear in $\size{\mathcal{X}}$ and it is 
independent of $\size{Q}$ and $\overline{\size{X}}$ in the ANNS. 
Although the query encoding is still linear in $\size{Q}$, the column 
embedding procedure can be accelerated by GPUs. 

% In the ranking stage, computing joinability for each candidate column runs 
% in $O(\size{Q} + \overline{\size{X}_C})$, where $\overline{\size{X}_C}$ 
% denotes the average size of the candidate columns. By assuming this value 
% is equal to the average over the entire repository, the time complexity of 
% the ranking stage is $O(c(\size{Q} + \overline{\size{X}}))$. 

% The overall time complexity of \deepjoin in online serving is 
% $O(\size{Q} + ... + vd\log\size{\mathcal{X}} + c(\size{Q} + \overline{\size{X}_C}))$. 
% Compared to \josie{}'s 
% $O(\size{\mathcal{X}} \cdot (\size{Q} + \overline{\size{X}}))$, we reduce 
% the times of joinability computation from $\size{\mathcal{X}}$ to $c$. 
% In addition, the search for candidates is regardless of the query column's 
% size because we embed it to a vector with a fixed dimensinality (768 for 
% the PLMs we use). In contrast, \josie needs to scan the prefix of the 
% column, whose length is often linear in the query column's size for set 
% similarity search, and access the corresponding postings lists in the 
% inverted index to find candidates. 

\section{Model Training}
\label{sec:train}
To fine-tune the PLM for joinability table discovery, 
we initialize the embedding model with the parameters of the PLM, and then 
train it with our training data and loss function. 
%Next, we introduce the our training data and loss function. 

\subsection{Training Data}
\label{sec:training-data}
Given a repository $\mathcal{X}$, we collect column pairs in $\mathcal{X}$ 
with high joinability as positive examples. This can be done by a 
self-join on $\mathcal{X}$ with a threshold $t$, i.e., finding column 
pairs $(X, Y)$ such that $X \in \mathcal{X}$, $Y \in \mathcal{X}$, and 
$jn(X, Y) \geq t$. To this end, we invoke a set similarity join~\cite{prefix-filter} 
for equi-joins or use \pexeso for semantic joins. In case $\mathcal{X}$ 
is large, we can perform the self-join on a sample of $\mathcal{X}$. 
% If the threshold $t$ is hard to specify beforehand, we may choose to find 
% \topk pairs of the set similarity join~\cite{probeset}, ranked by descending 
% order of joinability. 

In Definitions~\ref{def:equi-joinablity} and \ref{def:semantic-joinability}, 
the joinability is insensitive to the order of cells in a column, whereas PLMs 
are order-sensitive in their input. In order to make our model learn that the 
joinability is order-insensitive, we consider data augmentation by shuffling 
the cells in a column. In particular, we pick a percentage 
(called shuffle rate) of pairs $(X, Y)$ in the aforementioned positive 
examples, generate a random permutation of the cells of $X$, denoted by $X'$, 
and insert $(X', Y)$ to the set of positive examples. As such, the training 
set contains both $(X, Y)$ and $(X', Y)$, hence to suggest the 
order-insensitive joinability. Given a shuffle rate of $r$, out of all the 
positive examples, $r/(1 + r)$ of them are obtained from cell shuffle. 

To define negative training examples, we choose to use in-batch negatives, 
an easy and memory-efficient way that reuses the negative examples in the 
batch and has demonstrated effectiveness in text retrieval tasks~\cite{dpr}. 
Given a batch of positive training examples $\set{(X_i, Y_i)}$ (note that 
$X_i$ may be a shuffled column), we assume each $(X_i, Y_j), Y_i \neq Y_j$ 
as a negative pair. Despite a very small chance that $(X_i, Y_j)$ are 
joinable, this can be regarded as noise in the training data and our model 
is robust against this case. In our experiments, it shows better empirical 
results than other options of making negatives such as removing matching 
cells from positives. 

% Two possible data augmentation approaches: 
% \begin{itemize}
%     \item Positive: Shuffle the query column. There is no change in the 
%     set of joinable columns. 
%     \item Negative: Remove the cells that produce any join result. Then 
%     the joinable columns become unjoinable. This might be useful for 
%     sparse data where 1K out of 1M cells in the query column produce 
%     results, because the other cells may dominate the column embedding. 
% \end{itemize}

% \revise{
% For column-to-text transformation, e.g., for \texttt{col}, we may also 
% shuffle the column, concatenate the cells with ``,'' as delimiter, and 
% attach it to \texttt{col}. This may tell the PLM that the order of cells 
% in the column does not matter. 
% }

\subsection{Loss Function}
\label{sec:loss-function}
% Given a batch of $K$ training examples $\set{(X_i, Y_i)}$, where $X_i$ and 
% $Y_i$ are a pair of column embeddings, we assume each $(X_i, Y_i)$ is a 
% positive pair (i.e., the two columns are joinable) and each $(X_i, Y_j), 
% i \neq j$ is a negative pair. 
Given a batch of $N$ training examples $\set{(X_i, Y_i)}$, 
we minimize the multiple negative ranking loss~\cite{mnrloss}, which 
measures the negative log-likelihood of softmax normalized scores: 
\begin{align*}
    L(\vect{X}, \vect{Y}) &= -\frac{1}{N} \sum_{i=1}^{N} \log P_{\text{approx}} (Y_i \mid X_i) \\
                          &= -\frac{1}{N} \sum_{i=1}^{N} \left[ S(X_i, Y_i), - \log \sum_{j=1}^{N} \exp \left(S(X_i, Y_j) \right) \right].
\end{align*}

The above loss function is one of the prevalent options~\cite{sbert-loss} 
for fine-tuning sentence-transformers~\cite{sbert}. For the scoring function 
$S(\cdot, \cdot)$, we choose the cosine similarity of column embeddings, 
which shows the best empirical results. The subtlety here is that in the 
\topk retrieval, Euclidean distance is used instead for the ANNS. The choice 
of metrics will be evaluated in Section~\ref{sec:exp:ablation}. 
% We believe the difference is because that the length of embeddings is also 
% useful in identifying joinable results. 

\section{Experiments}
\label{sec:exp}
% \confversion{
% We report the most important experimental results here and provide the 
% following results in the extended version~\cite{extended-paper}: 
% \begin{inparaenum} [(1)]
%   \item accuracy and scalability with varying column size, 
%   \item ablation study for semantic joins, 
%   \item accuracy with varying vector matching threshold $\tau$ for 
%   semantic joins, 
%   \item more efficiency results (Wikitables and varying $k$), and 
%   \item more results for \deepjoindbert. 
% \end{inparaenum}
% In addition, since the effectiveness of joinable table discovery in data 
% analysis has been demonstrated in \cite{pexeso}, we do not repeat the 
% evaluation here.
% }

\begin{table} [t]
  \small
  \centering
  \caption{Dataset statistics.}  
  \resizebox{\linewidth}{!}{%
  \begin{tabular}{|l|l|c|c|c|c|c|} \hline
    Dataset & $\size{\mathcal{X}}$ & max. $\size{X}$ & min. $\size{X}$ & avg. $\size{X}$ & \# positive examples \\ \hline \hline
    Webtable-train  & 30k & 5454 & 5 & 20.77 & 190k (equi-), 220k (semantic) \\ \hline
    Wikitable-train & 30k & 1197 & 5 & 18.58 & 490k (equi-), 540k (semantic) \\ \hline
    Webtable-test   & 1M  & 6031 & 5 & 20.25 & N/A \\ \hline    
    Wikitable-test  & 1M  & 3454 & 5 & 18.71 & N/A \\ \hline  
  \end{tabular}
  }
  \label{tab:dataset}
\end{table}

\subsection{Experimental Settings}
\label{sec:exp:setup}

\begin{table*}[t]
  \centering
  \caption{Accuracy of equi-joins.}
  \scalebox{1}{
  \begin{tabular}{ l | ccccc| ccccc } 
    % \hline
    % \multicolumn{11}{c}{Webtable-small} \\ \hline
    % & \multicolumn{5}{c}{Precision@$k$} & \multicolumn{5}{c}{NDCG@$k$} \\ \hline
    % %Methods         & k = 10 &  Pre@20 &  Pre@30 &  Pre@40 & Pre@50  &  NDCG@10 &  NDCG@20 &  NDCG@30 &  NDCG@40 & NDCG@50 \\
    % Methods               & k = 10 &  20 &  30 &  40 & 50  & k = 10 &  20 &  30 &  40 & 50 \\
    % \hline
    % \lshensemble          & 0.574 & 0.563 & 0.554 & 0.545 & 0.535 & 0.752 & 0.761 & 0.766 & 0.772 & 0.777 \\
    % \fasttext             & 0.817 & 0.859 & 0.881 & 0.895 & 0.904 & 0.735 & 0.749 & 0.756 & 0.760 & 0.763 \\ 
    % \bert                 & 0.825 & 0.867 & 0.891 & 0.906 & 0.917 & 0.757 & 0.774 & 0.787 & 0.796 & 0.803 \\
    % \mpnet                & 0.826 & 0.869 & 0.895 & 0.911 & 0.920 & 0.755 & 0.771 & 0.786 & 0.797 & 0.806 \\
    % \tabert               & 0.790 & 0.841 & 0.870 & 0.888 & 0.900 & 0.696 & 0.723 & 0.738 & 0.750 & 0.761 \\
    % \mlp                  & 0.819 & 0.863 & 0.887 & 0.896 & 0.910 & 0.738 & 0.755 & 0.759 & 0.761 & 0.770 \\
    % \deepjoindbert (ours)  & 0.833 & 0.875 & 0.898 & 0.912 & 0.921 & 0.757 & 0.780 & 0.794 & 0.803 & 0.811 \\    
    % \deepjoinmpnet (ours)  & \textbf{0.868} & \textbf{0.905} & \textbf{0.925} & \textbf{0.937} & \textbf{0.944} & \textbf{0.814} & \textbf{0.832} & \textbf{0.843} & \textbf{0.851} & \textbf{0.857} \\
    % \hline

    \hline
    & \multicolumn{5}{c|}{Precision@$k$} & \multicolumn{5}{c}{NDCG@$k$} \\ \hline
    %Methods         & k = 10 &  Pre@20 &  Pre@30 &  Pre@40 & Pre@50  &  NDCG@10 &  NDCG@20 &  NDCG@30 &  NDCG@40 & NDCG@50 \\
    Methods         & $k$ = 10 &  20 &  30 &  40 & 50  & $k$ = 10 &  20 &  30 &  40 & 50 \\ \hline
    \multicolumn{11}{c}{Webtable} \\ \hline    
    \lshensemble          & 0.634 & 0.647 & 0.656 & 0.676 & 0.688 & 0.715 & 0.714 & 0.701 & 0.702 & 0.698 \\
    \fasttext             & 0.680 & 0.726 & 0.752 & 0.754 & 0.773 & 0.731 & 0.721 & 0.743 & 0.748 & 0.764 \\ 
    \bert                 & 0.652 & 0.695 & 0.712 & 0.722 & 0.729 & 0.698 & 0.713 & 0.708 & 0.707 & 0.708 \\  
    \mpnet                & 0.610 & 0.629 & 0.644 & 0.649 & 0.654 & 0.674 & 0.677 & 0.678 & 0.680 & 0.677 \\
    \tabert               & 0.622 & 0.637 & 0.645 & 0.656 & 0.671 & 0.694 & 0.685 & 0.690 & 0.693 & 0.691 \\ 
    \revise{\turl}        & 0.653 & 0.669 & 0.689 & 0.711 & 0.721 & 0.688 & 0.706 & 0.716 & 0.727 & 0.732 \\  
    \mlp                  & 0.683 & 0.719 & 0.755 & 0.758 & 0.778 & 0.737 & 0.735 & 0.748 & 0.755 & 0.769  \\  
    \deepjoindbert (ours) & 0.702 & 0.741 & 0.775 & 0.793 & 0.805 & 0.744 & 0.752 & 0.758 & 0.761 & 0.788 \\      
    \deepjoinmpnet (ours) & \textbf{0.732} & \textbf{0.775} & \textbf{0.791} & \textbf{0.812} & \textbf{0.832} & \textbf{0.768} & \textbf{0.786} & \textbf{0.799} & \textbf{0.803} & \textbf{0.822}  \\ 
    \hline

    % \hline
    % \multicolumn{11}{c}{Wikitable-small} \\ \hline
    % %& \multicolumn{5}{c}{Precision@$k$} & \multicolumn{5}{c|}{NDCG@$k$} \\ \hline
    % %Methods         & k = 10 &  Pre@20 &  Pre@30 &  Pre@40 & Pre@50  &  NDCG@10 &  NDCG@20 &  NDCG@30 &  NDCG@40 & NDCG@50 \\
    % %Methods         & k = 10 &  20 &  30 &  40 & 50  & k = 10 &  20 &  30 &  40 & 50 \\
    % \lshensemble          & 0.571 & 0.601 & 0.608 & 0.613 & 0.621 & 0.586 & 0.594 & 0.600 & 0.605 & 0.611 \\
    % \fasttext             & 0.663 & 0.733 & 0.767 & 0.787 & 0.805 & 0.657 & 0.677 & 0.686 & 0.700 & 0.710 \\ 
    % \bert                 & 0.640 & 0.718 & 0.759 & 0.785 & 0.807 & 0.658 & 0.681 & 0.698 & 0.715 & 0.727 \\  
    % \mpnet                & 0.647 & 0.717 & 0.759 & 0.784 & 0.804 & 0.657 & 0.677 & 0.690 & 0.701 & 0.711 \\
    % \tabert               & 0.645 & 0.703 & 0.751 & 0.776 & 0.795 & 0.633 & 0.645 & 0.658 & 0.697 & 0.699 \\  
    % \mlp                  & 0.671 & 0.744 & 0.771 & 0.793 & 0.810 & 0.671 & 0.682 & 0.693 & 0.705 & 0.717 \\  
    % \deepjoindbert (ours)   & 0.704 & 0.765 & 0.803 & 0.825 & 0.843 & 0.689 & 0.715 & 0.733 & 0.749 & 0.761 \\      
    % \deepjoinmpnet (ours)  & \textbf{0.723} & \textbf{0.778} & \textbf{0.819} & \textbf{0.844} & \textbf{0.865} & \textbf{0.702} & \textbf{0.733} & \textbf{0.741} & \textbf{0.767} & \textbf{0.793} \\   
    % \hline

    % \hline
    \multicolumn{11}{c}{Wikitable} \\ \hline
    %& \multicolumn{5}{c}{Precision@$k$} & \multicolumn{5}{c|}{NDCG@$k$} \\ \hline
    %Methods         & k = 10 &  Pre@20 &  Pre@30 &  Pre@40 & Pre@50  &  NDCG@10 &  NDCG@20 &  NDCG@30 &  NDCG@40 & NDCG@50 \\
    %Methods         & k = 10 &  20 &  30 &  40 & 50  & k = 10 &  20 &  30 &  40 & 50 \\
    \lshensemble          & 0.480 & 0.450 & 0.466 & 0.470 & 0.474 & 0.714 & 0.688 & 0.681 & 0.674 & 0.672 \\
    \fasttext             & 0.574 & 0.551 & 0.581 & 0.605 & 0.621 & 0.799 & 0.794 & 0.791 & 0.793 & 0.791 \\ 
    \bert                 & 0.436 & 0.460 & 0.497 & 0.520 & 0.541 & 0.719 & 0.721 & 0.731 & 0.736 & 0.740 \\  
    \mpnet                & 0.442 & 0.464 & 0.504 & 0.524 & 0.543 & 0.711 & 0.721 & 0.729 & 0.735 & 0.736 \\
    \tabert               & 0.431 & 0.445 & 0.488 & 0.520 & 0.539 & 0.701 & 0.708 & 0.732 & 0.725 & 0.737 \\ 
    \revise{\turl}        & 0.504 & 0.525 & 0.529 & 0.545 & 0.578 & 0.707 & 0.711 & 0.745 & 0.766 & 0.778 \\ 
    \mlp                  & 0.578 & 0.576 & 0.585 & 0.610 & 0.619 & 0.801 & 0.802 & 0.800 & 0.804 & 0.802 \\  
    \deepjoindbert (ours)   & 0.588 & 0.593 & 0.612 & 0.635 & 0.807 & 0.813 & 0.822 & 0.825 & 0.823 & 0.827 \\      
    \deepjoinmpnet (ours)  & \textbf{0.614} & \textbf{0.622} & \textbf{0.641} & \textbf{0.666} & \textbf{0.678} & \textbf{0.821} & \textbf{0.824} & \textbf{0.830} & \textbf{0.833} & \textbf{0.833} \\  
    \hline
    \end{tabular}
  }
  \label{tab:exp:equi-join}
\end{table*}

\begin{table*}[t]
  \centering
  \caption{Accuracy of semantic joins, $\tau = 0.9$ (labeled by \pexeso~\cite{pexeso}).}
  \scalebox{1}{
  \begin{tabular}{ l | ccccc|ccccc } 
    \hline
    & \multicolumn{5}{c|}{Precision@$k$} & \multicolumn{5}{c}{NDCG@$k$} \\ \hline
    Methods          & $k$ = 10 &  20 &  30 &  40 & 50  & $k$ = 10 &  20 &  30 &  40 & 50 \\ \hline
    \multicolumn{11}{c}{Webtable} \\ \hline
    %Methods         & k = 10 &  Pre@20 &  Pre@30 &  Pre@40 & Pre@50  &  NDCG@10 &  NDCG@20 &  NDCG@30 &  NDCG@40 & NDCG@50 \\
    \lshensemble     & 0.696 & 0.670 & 0.613 & 0.554 & 0.508 & 0.578 & 0.599 & 0.615 & 0.618 & 0.626 \\
    \fasttext        & 0.842 & 0.917 & 0.945 & 0.957 & 0.964 & 0.575 & 0.588 & 0.631 & 0.647 & 0.647 \\
    \deepjoindbert (ours) & 0.861 & 0.926 & 0.951 & 0.961 & 0.966 & 0.610 & 0.622 & 0.641 & 0.676 & 0.671 \\
    \deepjoinmpnet (ours)  & \textbf{0.874} & \textbf{0.934} & \textbf{0.954} & \textbf{0.963} & \textbf{0.970} & \textbf{0.640} & \textbf{0.657} & \textbf{0.664} & \textbf{0.685} & \textbf{0.680} \\ 
    \hline
    
    \multicolumn{11}{c}{Wikitable} \\ \hline
    %& \multicolumn{5}{c}{Precision@$k$} & \multicolumn{5}{c|}{NDCG@$k$} \\ \hline
    %Methods         & k = 10 &  Pre@20 &  Pre@30 &  Pre@40 & Pre@50  &  NDCG@10 &  NDCG@20 &  NDCG@30 &  NDCG@40 & NDCG@50 \\
    %Methods         & k = 10 &  20 &  30 &  40 & 50  & k = 10 &  20 &  30 &  40 & 50 \\
    %\hline
    \lshensemble     & 0.578 & 0.611 & 0.581 & 0.570 & 0.567 & 0.633 & 0.655 & 0.660 & 0.669 & 0.678 \\
    \fasttext        & 0.543 & 0.610 & 0.645 & 0.669 & 0.721 & 0.353 & 0.353 & 0.358 & 0.370 & 0.371 \\
    \deepjoindbert (ours) & 0.788 & 0.835 & 0.876 & 0.880 & 0.913 & 0.803 & 0.807 & 0.810 & 0.826 & 0.831 \\
    \deepjoinmpnet (ours)  & \textbf{0.813} & \textbf{0.881} & \textbf{0.889} & \textbf{0.889} & \textbf{0.936} & \textbf{0.814} & \textbf{0.820} & \textbf{0.833} & \textbf{0.842} & \textbf{0.852} \\ 
    \hline
    
    \end{tabular}
  }
  \label{tab:exp:pex:unsupervised}
\end{table*}

\begin{table*}[t]
  \centering
  \caption{Accuracy of semantic joins, $\tau = 0.8$ (labeled by \pexeso~\cite{pexeso}).}
  \scalebox{1}{
  \begin{tabular}{ l | ccccc|ccccc } 
    \hline
    & \multicolumn{5}{c|}{Precision@$k$} & \multicolumn{5}{c}{NDCG@$k$} \\ \hline
    \multicolumn{11}{c}{Webtable} \\ \hline
    %Methods         & k = 10 &  Pre@20 &  Pre@30 &  Pre@40 & Pre@50  &  NDCG@10 &  NDCG@20 &  NDCG@30 &  NDCG@40 & NDCG@50 \\
    Methods          & $k$ = 10 &  20 &  30 &  40 & 50  & $k$ = 10 &  20 &  30 &  40 & 50 \\
    \hline
    \lshensemble     & 0.571 & 0.592 & 0.621 & 0.613 & 0.633 & 0.604& 0.613 & 0.622 & 0.628 & 0.636 \\
    \fasttext        & 0.551 & 0.561 & 0.565 & 0.599 & 0.614 & 0.597 & 0.619 & 0.618 & 0.625 & 0.621 \\
    \deepjoindbert (ours) & 0.734 & 0.746 & 0.776 & 0.831 & 0.850 & 0.621 & 0.637 & 0.676 & 0.699 & 0.704 \\
    \deepjoinmpnet (ours)  & \textbf{0.774} & \textbf{0.791} & \textbf{0.823} & \textbf{0.845} & \textbf{0.881} & \textbf{0.655} & \textbf{0.684} & \textbf{0.723} & \textbf{0.729} & \textbf{0.737} \\ 
    \hline
    
    \multicolumn{11}{c}{Wikitable} \\ \hline
    %& \multicolumn{5}{c}{Precision@$k$} & \multicolumn{5}{c|}{NDCG@$k$} \\ \hline
    %Methods         & k = 10 &  Pre@20 &  Pre@30 &  Pre@40 & Pre@50  &  NDCG@10 &  NDCG@20 &  NDCG@30 &  NDCG@40 & NDCG@50 \\
    %Methods         & k = 10 &  20 &  30 &  40 & 50  & k = 10 &  20 &  30 &  40 & 50 \\
    %\hline
    \lshensemble     & 0.499 & 0.529 & 0.497 & 0.491 & 0.504 & 0.573 & 0.570 & 0.569 & 0.573 & 0.582 \\
    \fasttext        & 0.395 & 0.480 & 0.523 & 0.549 & 0.607 & 0.203 & 0.204 & 0.210 & 0.222 & 0.223 \\
    \deepjoindbert (ours) & 0.621 & 0.714 & 0.758 & 0.776 & 0.811 & 0.598 & 0.632 & 0.676& 0.688 & 0.703 \\
    \deepjoinmpnet (ours)  & \textbf{0.659} & \textbf{0.758} & \textbf{0.803} & \textbf{0.805} & \textbf{0.846} & \textbf{0.620} & \textbf{0.670} & \textbf{0.694} & \textbf{0.710} & \textbf{0.722} \\ 
    \hline
    
    \end{tabular}
  }
  \label{tab:exp:pex:unsupervised:tau0.8}
\end{table*}

\begin{table*}[t]
  \centering
  \caption{Accuracy of semantic joins, $\tau = 0.7$ (labeled by \pexeso~\cite{pexeso}).}
  \scalebox{1}{
  \begin{tabular}{ l | ccccc|ccccc } 
    \hline
    & \multicolumn{5}{c|}{Precision@$k$} & \multicolumn{5}{c}{NDCG@$k$} \\ \hline
    Methods          & $k$ = 10 &  20 &  30 &  40 & 50 & $k$ = 10 &  20 &  30 &  40 & 50 \\ \hline
    \multicolumn{11}{c}{Webtable} \\ \hline
    \lshensemble     & 0.321 & 0.368 & 0.389 & 0.394 & 0.390 & 0.333 & 0.338 & 0.355 & 0.364 & 0.377 \\
    \fasttext        & 0.397 & 0.505 & 0.594 & 0.663 & 0.722 & 0.352 & 0.370 & 0.384 & 0.422 & 0.440 \\
    \deepjoindbert (ours) & 0.411 & 0.509 & 0.601 & 0.673 & 0.738 & 0.359 & 0.381 & 0.396 & 0.433 & 0.461 \\
    \deepjoinmpnet (ours)  & \textbf{0.426} & \textbf{0.527} & \textbf{0.604} & \textbf{0.679} & \textbf{0.742} & \textbf{0.363} & \textbf{0.388} & \textbf{0.411} & \textbf{0.435} & \textbf{0.471} \\ 
    \hline
    
    \multicolumn{11}{c}{Wikitable} \\ \hline
    \lshensemble     & 0.310 & 0.346 & 0.336 & 0.351 & 0.342 & 0.474 & 0.477 & 0.470 & 0.467 & 0.470 \\
    \fasttext        & 0.093 & 0.140 & 0.190 & 0.230 & 0.256 & 0.058 & 0.067 & 0.075 & 0.082 & 0.086 \\
    \deepjoindbert (ours) & 0.431 & 0.497 & 0.523 & 0.554 & 0.575 & 0.601 & 0.607 & 0.611 & 0.626 & 0.624 \\
    \deepjoinmpnet (ours)  & \textbf{0.476} & \textbf{0.539} & \textbf{0.568} & \textbf{0.593} & \textbf{0.604} & \textbf{0.623} & \textbf{0.627} & \textbf{0.631} & \textbf{0.646} & \textbf{0.647} \\ 
    \hline
    
    \end{tabular}
  }
  \label{tab:exp:pex:unsupervised:tau0.7}
\end{table*}

\begin{table}[t]
  \centering
  \caption{Accuracy of semantic joins, $\tau = 0.9$ (labeled by experts).}
  \scalebox{1}{
  \begin{tabular}{ l | c | c | c } 
    \hline
    %Methods         & k = 10 &  Pre@20 &  Pre@30 &  Pre@40 & Pre@50  &  NDCG@10 &  NDCG@20 &  NDCG@30 &  NDCG@40 & NDCG@50 \\
    Methods          & Precision & Recall & F1 \\ \hline
    \multicolumn{4}{c}{Webtable} \\ \hline    
    \lshensemble     & 0.181 & 0.228 & 0.202   \\
    \fasttext        & 0.138 & 0.277 & 0.183   \\
    \pexeso          & 0.212 & 0.506 & 0.300   \\
    \deepjoinmpnet (ours)  & \textbf{0.350}  & \textbf{0.693} & \textbf{0.465}  \\ 
    \hline
    
    \multicolumn{4}{c}{Wikitable}\\ \hline
    \lshensemble     & 0.652 & 0.385 & 0.484  \\
    \fasttext        & 0.467 & 0.380 & 0.419  \\
    \pexeso          & 0.683 & 0.492 & 0.572  \\
    \deepjoinmpnet (ours)  & \textbf{0.842} & \textbf{0.568} & \textbf{0.677} \\ 
    \hline
    \end{tabular}
  }
  \label{tab:exp:pex:human}
\end{table}

\begin{table*} [t]
  \small
  \centering
  \caption{\revise{Win/lose examples for \deepjoinmpnet v.s. non-\deepjoin competitors on semantic joins labeled by experts.}}  
  \resizebox{\linewidth}{!}{%
  \begin{tabular}{|l|l|c|l|} \hline
    Query & \deepjoinmpnet's result & Win/Lose & Possible reason \\ \hline 
    \texttt{title}: Headboard Buying Guide & \texttt{title}: Carved Headboard | west elm & & The attention mechanism in \deepjoinmpnet \\ 
    \texttt{colname}: Mattress Size & \texttt{colname}: Item & Win & focuses on the word ``Headboard'' and \\
    \texttt{col}: California King, Full/Double, King, Queen, Twin, ... & \texttt{col}: Carved Headboard Full, Carved Headboard King, ... & & bed size words such as ``King'' and ``Double''. \\ \hline 
    \texttt{title}: Tallest buildings & \texttt{title}: Buildings above 140m & & The PLM in \deepjoinmpnet captures the semantic similarity \\ 
    \texttt{colname}: Name & \texttt{colname}: Name & Win & between ``Tallest buildings'' and ``Buildings above 140m'' in titles, \\
    \texttt{col}: City Tower, City-Haus, City-Hochhaus Leipzig, ... & \texttt{col}: Centrum LIM, City-Haus, City-Hochhaus, ... & &  as well as tall building names in cell values.\\ \hline 
    \texttt{title}: How to call Kazakhstan from Korea South & \texttt{title}: How to call Georgia from Georgia & & The PLM in \deepjoinmpnet pays too much attention \\ 
    \texttt{colname}: City & \texttt{colname}: City & Lose & to metadata, but ignores that there are \\
    \texttt{col}: Aktubinsk, Almaty, Arkalyk, ... & \texttt{col}: Akhalgori, Akhmeta, Aspindza, ... & & no similar values between the column contents.\\ \hline  
  \end{tabular}
  }
  \label{tab:win-lose-example}
\end{table*}

\begin{table} [t]
  \small
  \centering
  \caption{\revise{Recall@10, semantic joins drill-down: near duplicates (\textsf{ND}) and attribute enrichment (\textsf{AE}).}}
  %\resizebox{\linewidth}{!}{%
  \begin{tabular}{l | cc|cc} \hline
    & \multicolumn{2}{c|}{Webtable} & \multicolumn{2}{c}{Wikitable}\\ \hline
    Methods        & \textsf{ND} & \textsf{AE} & \textsf{ND} & \textsf{AE} \\ \hline 
    \lshensemble   & 0.320 & 0.381 & 0.344 & 0.274 \\
    \fasttext      & 0.195 & 0.262 & 0.373 & 0.334 \\ 
    \pexeso        & 0.355 & 0.399 & 0.215 & 0.335 \\ 
    \deepjoinmpnet & \textbf{0.701} & \textbf{0.561} & \textbf{0.649} & \textbf{0.418} \\ \hline
  \end{tabular}
  %}
  \label{tab:join-case-drill-down}
\end{table}

\begin{table}[t]
  \centering
  \caption{Effect of varying column size.}
  \resizebox{\linewidth}{!}{
  \begin{tabular}{ l | ccc| ccc } 
    \hline
    & \multicolumn{3}{c}{Precision@10} & \multicolumn{3}{c}{NDCG@10} \\ \hline
    Methods                & $\size{X}$ = 5 -- 10 &  10 -- 50 & $>$ 50   & $\size{X}$ = 5 -- 10 &  10 -- 50 & $>$ 50  \\ \hline
    \multicolumn{7}{c}{Webtable, equi-joins} \\ \hline
    %Methods         & k = 10 &  Pre@20 &  Pre@30 &  Pre@40 & Pre@50  &  NDCG@10 &  NDCG@20 &  NDCG@30 &  NDCG@40 & NDCG@50 \\
    \lshensemble           & 0.647 & 0.633 & 0.617 & 0.722 & 0.693 & 0.688 \\
    \fasttext              & 0.692 & 0.694 & 0.673 & 0.764 & 0.751 & 0.719 \\ 
    \bert                  & 0.684 & 0.663 & 0.642 & 0.755 & 0.731 & 0.714 \\  
    \mpnet                 & 0.627 & 0.619 & 0.614 & 0.718 & 0.698 & 0.699 \\
    \tabert                & 0.652 & 0.651 & 0.649 & 0.724 & 0.731 & 0.702 \\ 
    \revise{\turl}         & 0.678 & 0.667 & 0.645 & 0.729 & 0.744 & 0.715  \\ 
    \mlp                   & 0.695 & 0.691 & 0.664 & 0.765 & 0.755 & 0.701 \\  
    \deepjoindbert (ours)  & 0.724 & 0.711 & 0.703 & 0.777 & 0.768 & 0.761 \\      
    \deepjoinmpnet (ours)  & \textbf{0.765} & \textbf{0.741} & \textbf{0.737} & \textbf{0.789} & \textbf{0.773} & \textbf{0.764} \\ 
    \hline

    %\hline
    \multicolumn{7}{c}{Webtable, semantic joins} \\ \hline
%    & \multicolumn{3}{c}{Precision@$k$} & \multicolumn{3}{c}{NDCG@$k$} \\ \hline    
    %Methods                & $\size{X}$ = 5 -- 10 &  10 -- 50 & $>$ 50   & $\size{X}$ = 5 -- 10 &  10 -- 50 & $>$ 50  \\    
    \lshensemble           & 0.722 & 0.721 & 0.714 & 0.621 & 0.618 & 0.605 \\
    \fasttext              & 0.851 & 0.841 & 0.837 & 0.613 & 0.622 & 0.616 \\
    \deepjoindbert (ours)  & 0.878 & 0.851 & 0.849 & 0.645 & 0.640 & 0.638 \\
    \deepjoinmpnet (ours)  & \textbf{0.884} & \textbf{0.871} & \textbf{0.856} & \textbf{0.677} & \textbf{0.655} & \textbf{0.651} \\ 
    \hline
    \end{tabular}
  }
  \label{tab:exp:accuracy-column-size}
\end{table}

\begin{table}[t]
  \centering
  \caption{\revise{Evaluation on tall columns, synthetic, equi-joins.}}
  \resizebox{\linewidth}{!}{
  \begin{tabular}{ l | cccc| cccc } 
    \hline
    & \multicolumn{4}{c}{Precision@10} & \multicolumn{4}{c}{NDCG@10} \\ \hline
    %Methods         & k = 10 &  Pre@20 &  Pre@30 &  Pre@40 & Pre@50  &  NDCG@10 &  NDCG@20 &  NDCG@30 &  NDCG@40 & NDCG@50 \\
    Methods                                  & $\size{X}$ = 512 & 1024 & 2048 & 4096  & $\size{X}$ = 512 & 1024 & 2048 & 4096  \\ \hline
    \lshensemble                             & 0.518 & 0.521 & 0.505& 0.511 & 0.611 & 0.606 & 0.597 & 0.593\\
    \fasttext                                & 0.577 & 0.568 & 0.561 & 0.563 & 0.634 & 0.639 & 0.624 & 0.615\\ 
    \bert                                    & 0.578 & 0.554 & 0.545 & 0.547 & 0.635 & 0.622 & 0.621 & 0.600\\  
    \mpnet                                   & 0.545 & 0.531 & 0.537 & 0.534 & 0.622 & 0.615 & 0.611 & 0.601\\
    \tabert-\textsf{random}                  & 0.523 & 0.521 & 0.517 & 0.529 & 0.619 & 0.610 & 0.606 & 0.592\\ 
    \turl-\textsf{random}                    & 0.534 & 0.529 & 0.511 & 0.521 & 0.627 & 0.612 & 0.601 & 0.587\\ 
    \mlp                                     & 0.581 & 0.569 & 0.563 & 0.571 & 0.639 & 0.638 & 0.626 & 0.617\\  
    \deepjoindbert-\textsf{frequency} (ours) & 0.664 & 0.647 & 0.622 & 0.617 & 0.677 & 0.668 & 0.657 & 0.644\\ %\hline
    \deepjoinmpnet-\textsf{frequency} (ours) & \textbf{0.697} & \textbf{0.671} & \textbf{0.666} & \textbf{0.669} & \textbf{0.696} & \textbf{0.674} & \textbf{0.677} & \textbf{0.665} \\ 
    \deepjoinmpnet-\textsf{random} (ours)    & 0.684 & 0.669 & 0.643 & 0.641 & 0.691 & \textbf{0.674} & 0.664 & 0.649 \\ 
    \deepjoinmpnet-\textsf{truncate} (ours)  & 0.681 & 0.657 & 0.645 & 0.642 & 0.683 & 0.655 & 0.661 & 0.654 \\ 
    \hline
    \end{tabular}
    }
  \label{tab:exp:accuracy-tall-columns}
\end{table}

\mysubsection{Datasets}
The following two datasets are used in the evaluation. 
\begin{inparaenum} [(1)]
  \item \textbf{Webtable} is a dataset of the WDC Web Table Corpus~\cite{WDC}. 
  We use the English relational web tables 2015 and for each table, we 
  extract the key column defined in the metadata. 
  \item \textbf{Wikitable} is a dataset of relational tables from Wikipedia~\cite{wikitable}. 
  For each table, we take the column which contains the largest number 
  of distinct values in the table. 
\end{inparaenum} 
Both datasets contain metadata for table title, column name, and context, 
and have been used in previous 
works~\cite{josie, pexeso, lsh-ensemble, tabert, gtr, tabel}. Columns 
that are too short ($< 5$ cells) are removed. For semantic joins, 
fastText~\cite{fasttext} is used to embed cells, Euclidean distance is 
used for distance function $d$, and the threshold $\tau$ for vector 
matching is 
%\confversion{0.9.} 
0.9, unless otherwise specified. 

%\mysubsection{Training and testing data}
In order to show that \deepjoin learned from a small subset of a corpus 
is able to generalize to a large subset, we randomly sample two subsets 
of 30k and 1M columns, respectively, from each corpus. From the 30k 
training set, we randomly sample column pairs whose $jn \geq 0.7$ as 
initial positive examples, where $jn$ is defined using 
Equation~\ref{eq:equi-joinability} for equi-joins or 
Equation~\ref{eq:semantic-joinability} for semantic joins. We then apply 
the techniques in Section~\ref{sec:training-data} for data augmentation 
and making negative examples. The 1M testing set is used as the repository 
$\mathcal{X}$ for search. To generate queries and avoid data leaks, we 
randomly sample 50 columns from the original corpus except those in 
$\mathcal{X}$. The dataset statistics are given in Table~\ref{tab:dataset}.

\mysubsection{Methods}
We compare the following methods. 
\begin{inparaenum} [(1)]
    \item \deepjoin: This is our proposed model. We equip our model with 
    DistilBERT~\cite{distilbert} and MPNet~\cite{mpnet} as PLM and denote 
    the resultant model as \deepjoindbert and \deepjoinmpnet, respectively. 
    \item \josie~\cite{josie}: This is an exact solution to equi-joinable 
    table discovery, based on \topk set similarity search.
    \item \lshensemble~\cite{lsh-ensemble}: This is an approximate solution 
    to equi-joinable table discovery, based on partitioning and MinHash.
    \item \fasttext, \bert, \mpnet: We replace the column embedding in 
    \deepjoin by averaging (no fine-tuning) the word embeddings from 
    fastText~\cite{fasttext}, BERT~\cite{bert}, and MPNet~\cite{mpnet}, 
    respectively. 
    \item \tabert~\cite{tabert}: This is a table embedding approach which 
    uses BERT and learns column embeddings for question answering tasks. We 
    use its column embedding to replace that in \deepjoin. 
    \item \revise{\turl~\cite{turl}: This is a representation learning approach 
    for table understanding tasks. We use its column embedding to replace that in 
    \deepjoin.} 
    \item \mlp: We replace the PLM in \deepjoin with a 3-layer perceptron 
    trained for a regression, which takes as input the fastText 
    embeddings of two columns and outputs the joinability. Then, we take 
    the output of the last hidden layer as column embedding. 
    \item \pexeso~\cite{pexeso}: This is an exact solution to 
    semantic-joinable table discovery, using pivot-based filtering and a 
    grid index. 
    % It converts cell values to vectors by \fasttext and defines matching 
    % cells as vectors whose distance is no larger than a threshold $t$. Then, 
    % it uses a pivot-based index to find columns with at least $T$ matching 
    % cells to the query column. 
\end{inparaenum}

\mysubsection{Metrics}
For accuracy, we evaluate precision@$k$ and normalized discounted 
cumulative gain (NDCG@$k$). Precision@$k$ measures the overlap 
between the model's \topk results and the \topk of an 
exact solution to Problem~\ref{pb:joinable-table-discovery}. NDCG@$k$ is 
defined as 
$\frac{DCG_{\text{model}}}{DCG_{\text{exact}}}$, where 
$DCG = \sum_{i=1}^{k} \frac{jn(Q, X_i)}{\log_2 (i + 1)}$, and the $X_i$'s for 
$DCG_{\text{model}}$ and $DCG_{\text{exact}}$ are the \topk of the model and the exact solution, respectively. 
For semantic joins, we also request our 
colleagues of database researchers to label whether a retrieved table is 
really joinable, and then measure precision, recall, and F1 score. 
Precision $=$ (\# retrieved joinable columns) $/$ (\# retrieved columns). 
Since it is too laborious to label every table in the dataset, we follow 
\cite{recall-search-engine} and build a retrieved pool using the union of the 
tables identified by the compared methods, \revise{which is also common practice for 
the evaluation of Web search engines.} 
Recall $=$ (\# retrieved joinable columns) $/$ (\# joinable columns in the 
retrieved pool), \revise{where joinable columns are labeled by our experts.} 
For efficiency, we evaluate the end-to-end processing time, including 
column-to-text transformation, query embedding, and ANNS. 
The above measures are averaged over all the queries. 
% Accuracy and efficiency are reported by averaging over all the queries 
% used in the experiments.

\mysubsection{Environments} 
\deepjoin are implemented with PyTorch. \revise{
we use the Sentence-BERT~\cite{sbert-url} 
and the Hugging Face~\cite{hugging-face-transformers} 
libraries to build and train the \deepjoin model.} 
We use the following setting: batch size = 32, 
learning rate = 2e-5, warmup steps = 10000, and weight decay rate = 0.01. 
Like \deepjoin, other column embedding methods (\fasttext, \bert, 
\mpnet, \tabert, \turl, and \mlp) follow the same ANNS scheme, for which we 
use IVFPQ~\cite{ivfpq} and HNSW~\cite{hnsw} in the Faiss library~\cite{faiss}. 
Experiments are run on 
a server with a 2.20GHz Intel Xeon CPU E7-8890 and 630 GB RAM. Models are 
(optionally) accelerated using a NVidia A100 Tensor Core. All the 
competitors are implemented in Python 3.7.

\begin{table*}[t]
  \centering
  \caption{Evaluation of column-to-text transformation, equi-joins.}
  \scalebox{1}{
  \begin{tabular}{ l | ccccc| ccccc  } 
    \hline
    & \multicolumn{5}{c|}{Precision@$k$} & \multicolumn{5}{c}{NDCG@$k$} \\ \hline
    Methods         & $k$ = 10 &  20 &  30 &  40 & 50  & $k$ = 10 &  20 &  30 &  40 & 50 \\
    \hline
    \multicolumn{11}{c}{Webtable} \\ \hline
    %Methods         & k = 10 &  Pre@20 &  Pre@30 &  Pre@40 & Pre@50  &  NDCG@10 &  NDCG@20 &  NDCG@30 &  NDCG@40 & NDCG@50 \\
    \texttt{col}                       & 0.700 & 0.744 & 0.763 & 0.788 & 0.791 & 0.745 & 0.753 & 0.767 & 0.779 & 0.795 \\
    \texttt{colname-col}               & 0.709 & 0.750 & 0.771 & 0.795 & 0.799 & 0.751 & 0.757 & 0.770 & 0.785 & 0.802 \\ 
    \texttt{colname-col-context}       & 0.703 & 0.746 & 0.764 & 0.795 & 0.798 & 0.750 & 0.755 & 0.770 & 0.780 & 0.800 \\
    \texttt{colname-stat-col}          & 0.712 & 0.757 & 0.778 & 0.799 & 0.799 & 0.756 & 0.758 & 0.773 & 0.788 & 0.805 \\  
    \texttt{title-colname-col}         & 0.729 & 0.771 & 0.785 & 0.807 & 0.821 & 0.761 & 0.769 & 0.788 & 0.795 & 0.818 \\
    \texttt{title-colname-col-context} & 0.718 & 0.759 & 0.781 & 0.799 & 0.820 & 0.759 & 0.766 & 0.784 & 0.791 & 0.815 \\ 
    \texttt{title-colname-stat-col}    & \textbf{0.732} & \textbf{0.775} & \textbf{0.791} & \textbf{0.812} & \textbf{0.832} & \textbf{0.768} & \textbf{0.786} & \textbf{0.799} & \textbf{0.803} & \textbf{0.822}  \\ 
    \hline
    
    \multicolumn{11}{c}{Wikitable} \\ 
    \hline
    \texttt{col}                       & 0.602 & 0.604 & 0.617 & 0.632 & 0.651 & 0.804 & 0.805 & 0.812 & 0.819 & 0.821 \\
    \texttt{colname-col}               & 0.600 & 0.607 & 0.615 & 0.630 & 0.654 & 0.801 & 0.816 & 0.817 & 0.821 & 0.822\\ 
    \texttt{colname-col-context}       & 0.599 & 0.607 & 0.613 & 0.628 & 0.655 & 0.805 & 0.814 & 0.818 & 0.819 & 0.821\\
    \texttt{colname-stat-col}          & 0.605 & 0.608 & 0.617 & 0.635 & 0.663 & 0.801 & 0.814 & 0.815 & 0.822  & 0.824\\  
    \texttt{title-colname-col}         & 0.611 & 0.614 & 0.627 & 0.647 & 0.671 & 0.813 & 0.820 & 0.824 & 0.827 & \textbf{0.833}\\
    \texttt{title-colname-col-context} & 0.608 & 0.618 & 0.630 & 0.644 & 0.670 & 0.815 & 0.821 & 0.822 & 0.828 & 0.831\\ 
    \texttt{title-colname-stat-col}    & \textbf{0.614} & \textbf{0.622} & \textbf{0.641} & \textbf{0.666} & \textbf{0.678} & \textbf{0.821} & \textbf{0.824} & \textbf{0.830} & \textbf{0.833} & \textbf{0.833} \\  
    \hline
    \end{tabular}
  }
  \label{tab:exp:col2text:equi}
\end{table*}

\begin{table*}[t]
  \centering
  \caption{Evaluation of column-to-text transformation, semantic joins.}
  \scalebox{1}{
  \begin{tabular}{ l | ccccc| ccccc  } 
    \hline
    & \multicolumn{5}{c|}{Precision@$k$} & \multicolumn{5}{c}{NDCG@$k$} \\ \hline
    Methods         & $k$ = 10 &  20 &  30 &  40 & 50  & $k$ = 10 &  20 &  30 &  40 & 50 \\
    \hline
    \multicolumn{11}{c}{Webtable} \\ \hline
    %Methods         & k = 10 &  Pre@20 &  Pre@30 &  Pre@40 & Pre@50  &  NDCG@10 &  NDCG@20 &  NDCG@30 &  NDCG@40 & NDCG@50 \\
    \texttt{col}                       & 0.826 & 0.833 & 0.866 & 0.885 & 0.925 & 0.610 & 0.615 & 0.623 & 0.637 & 0.644 \\
    \texttt{colname-col}               & 0.831 & 0.840 & 0.877 & 0.899 & 0.945 & 0.616 & 0.620 & 0.631 & 0.644 & 0.652 \\ 
    \texttt{colname-col-context}       & 0.831 & 0.839 & 0.875 & 0.886 & 0.945 & 0.620 & 0.631 & 0.640 & 0.650 & 0.661\\
    \texttt{colname-stat-col}          & 0.834 & 0.846 & 0.887 & 0.904 & 0.956 & 0.625 & 0.641 & 0.659 & 0.654 & 0.671\\  
    \texttt{title-colname-col}         & 0.851 & 0.879 & 0.904 & 0.926 & 0.959 & 0.633 & 0.651 & 0.667 & 0.670 & 0.675 \\
    \texttt{title-colname-col-context} & 0.850 & 0.877 & 0.915 & 0.927 & 0.954 & 0.631 & 0.650 & 0.671 & 0.675 & 0.677 \\ 
    \texttt{title-colname-stat-col}    & \textbf{0.874} & \textbf{0.934} & \textbf{0.954} & \textbf{0.963} & \textbf{0.970} & \textbf{0.640} & \textbf{0.657} & \textbf{0.664} & \textbf{0.685} & \textbf{0.680}  \\ 
    \hline
    
    \multicolumn{11}{c}{Wikitable} \\ 
    \hline
    \texttt{col}                       & 0.773 & 0.810 & 0.837 & 0.845 & 0.891 & 0.791 & 0.803 & 0.807 & 0.822 & 0.825  \\
    \texttt{colname-col}               & 0.775 & 0.815 & 0.842 & 0.847 & 0.903 & 0.797 & 0.807 & 0.811 & 0.829 & 0.834 \\ 
    \texttt{colname-col-context}       & 0.774 & 0.812 & 0.841 & 0.847 & 0.901 & 0.794 & 0.807 & 0.810 & 0.830 & 0.833 \\
    \texttt{colname-stat-col}          & 0.784 & 0.820 & 0.850 & 0.855 & 0.913 & 0.804 & 0.811 & 0.815 & 0.833 & 0.841 \\  
    \texttt{title-colname-col}         & 0.804 & 0.836 & 0.868 & 0.874 & 0.922 & 0.811 & 0.815 & 0.821 & 0.837 & 0.844 \\
    \texttt{title-colname-col-context} & 0.803 & 0.835 & 0.868 & 0.877 & 0.923 & 0.811 & 0.817 & 0.826 & 0.840 & 0.845 \\ 
    \texttt{title-colname-stat-col}    & \textbf{0.813} & \textbf{0.881} & \textbf{0.889} & \textbf{0.889} & \textbf{0.936} & \textbf{0.814} & \textbf{0.820} & \textbf{0.833} & \textbf{0.842} & \textbf{0.852} \\  
    \hline
    \end{tabular}
  }
  \label{tab:exp:col2text:semantic}
\end{table*}

\begin{table*}[t]
  \centering
  \caption{Evaluation of cell shuffle, equi-joins.}
  \scalebox{1}{
  \begin{tabular}{ c | ccccc|ccccc } 
    \hline
    & \multicolumn{5}{c|}{Precision@$k$} & \multicolumn{5}{c}{NDCG@$k$} \\ \hline
    shuffle rate         & $k$ = 10 &  20 &  30 &  40 & 50  & $k$ = 10 &  20 &  30 &  40 & 50 \\
    \hline
    \multicolumn{11}{c}{Webtable} \\ \hline
    %Methods         & k = 10 &  Pre@20 &  Pre@30 &  Pre@40 & Pre@50  &  NDCG@10 &  NDCG@20 &  NDCG@30 &  NDCG@40 & NDCG@50 \\
    \texttt{0.0}     & 0.720 & 0.759 & 0.781 & 0.803 & 0.819 & 0.752 & 0.771 & 0.784 & 0.791 & 0.812 \\
    \texttt{0.1}            & 0.725 & 0.766 & 0.784 & 0.809 & 0.825 & 0.755 & 0.778 & 0.793 & 0.796 & 0.817 \\
    \texttt{0.2}            & \textbf{0.732} & \textbf{0.775} & \textbf{0.791} & \textbf{0.812} & \textbf{0.832} & \textbf{0.768} & \textbf{0.786} & \textbf{0.799} & \textbf{0.803} & \textbf{0.822}  \\  
    \texttt{0.3}                 & 0.729 & 0.770 & 0.785 & 0.792 & 0.815 & 0.754 & 0.773 & 0.788 & 0.791 & 0.806 \\
    \texttt{0.4}                 & 0.711 & 0.755 & 0.774 & 0.780 & 0.782 & 0.733 & 0.758 & 0.766 & 0.780 & 0.781\\  
    \texttt{0.5}                 & 0.701 & 0.751 & 0.760 & 0.781 & 0.787 & 0.726 & 0.754 & 0.760 & 0.765 & 0.777\\
    \hline
    
    \multicolumn{11}{c}{Wikitable} \\ \hline
    \texttt{0.0} & 0.605 & 0.615 & 0.631 & 0.657 & 0.670 & 0.811 & 0.813 & 0.815 & 0.826 & 0.821 \\
    \texttt{0.1}            & 0.608 & 0.618 & 0.635 & 659 & 0.675 & 0.809 & 0.814 & 0.829 & 0.828 & 0.829 \\
    \texttt{0.2}            & 0.611 & \textbf{0.622} & \textbf{0.664} &0.639 & 0.677 & 0.815 & 0.820 & 0.831 & 0.832 & 0.830 \\ 
    \texttt{0.3}            & \textbf{0.614} & \textbf{0.622} & 0.641 & \textbf{0.666} & \textbf{0.678} & \textbf{0.821} & \textbf{0.824} & \textbf{0.830} & \textbf{0.833} & \textbf{0.833} \\
    \texttt{0.4}            & 0.584 & 0.598 & 0.613 & 0.634 & 0.644 & 0.803 & 0.801 & 0.813 & 0.815 & 0.821 \\  
    \texttt{0.5}            & 0.576 & 0.579 & 0.591 & 0.623 & 0.634 & 0.800 & 0.797 & 0.802 & 0.808 & 0.810 \\
    \hline
    \end{tabular}
  }
  \label{tab:exp:data-aug:equi}
\end{table*}

\begin{table*}[t]
  \centering
  \caption{Evaluation of cell shuffle, semantic joins.}
  \scalebox{1}{
  \begin{tabular}{ c | ccccc|ccccc } 
    \hline
    & \multicolumn{5}{c|}{Precision@$k$} & \multicolumn{5}{c}{NDCG@$k$} \\ \hline
    shuffle rate         & $k$ = 10 &  20 &  30 &  40 & 50  & $k$ = 10 &  20 &  30 &  40 & 50 \\
    \hline
    \multicolumn{11}{c}{Webtable} \\ \hline
    %Methods         & k = 10 &  Pre@20 &  Pre@30 &  Pre@40 & Pre@50  &  NDCG@10 &  NDCG@20 &  NDCG@30 &  NDCG@40 & NDCG@50 \\
    \texttt{0.0}     & 0.868 & 0.917 & 0.950 & 0.954 & 0.959 & 0.631 & 0.649 & 0.651 & 0.677 & 0.679 \\
    \texttt{0.1}            & 0.870 & 0.919 & 0.949 & 0.959 & 0.963 & 0.633 & 0.651 & 0.655 & 0.679 & 0.683 \\
    \texttt{0.2}            & 0.872 & 0.922 & 0.950 & 0.961 & 0.966 & 0.639 & 0.655 & 0.659 & 0.681 & \textbf{0.687}  \\  
    \texttt{0.3}            & \textbf{0.874} & \textbf{0.934} & \textbf{0.954} & \textbf{0.963} & \textbf{0.970} & \textbf{0.640} & \textbf{0.657} & \textbf{0.664} & \textbf{0.685} & 0.680  \\ 
    \texttt{0.4}            & 0.871 & 0.930 & 0.939 & 0.961 & 0.968 & 0.631 & 0.654 & 0.654 & 0.683 & 0.686\\  
    \texttt{0.5}            & 0.863 & 0.919 & 0.945 & 0.955 & 0.961 & 0.632 & 0.649 & 0.648 & 0.679 & 0.681\\
    \hline
    
    \multicolumn{11}{c}{Wikitable} \\ \hline
    \texttt{0.0}     & 0.798 & 0.856 & 0.865 & 0.877 & 0.914 & 0.801 & 0.804 & 0.813 & 0.820 & 0.833 \\
    \texttt{0.1}            & 0.801 & 0.861 & 0.870 & 0.881 & 0.921 & 0.803 & 0.806 & 0.819 & 0.822 & 0.839 \\
    \texttt{0.2}            & 0.806 & 0.866 & 0.875 & 0.883 & 0.925 & 0.806 & 0.810 & 0.822 & 0.825 & 0.840 \\ 
    \texttt{0.3}            & 0.808 & 0.870 & 0.877 & 0.887 & 0.929 & 0.809 & 0.812 & 0.825 & 0.826 & 0.843 \\
    \texttt{0.4}            & \textbf{0.813} & \textbf{0.881} & \textbf{0.889} & \textbf{0.889} & \textbf{0.936} & \textbf{0.814} & \textbf{0.820} & \textbf{0.833} & \textbf{0.842} & \textbf{0.852}  \\  
    \texttt{0.5}            & 0.809 & 0.871 & 0.873 & 0.880 & 0.931 & 0.809 & 0.813 & 0.829 & 0.829 & 0.844 \\
    \hline
    \end{tabular}
  }
  \label{tab:exp:data-aug:semantic}
\end{table*}

\subsection{Accuracy Evaluation}
\label{sec:exp:accuracy}
For equi-join, Table~\ref{tab:exp:equi-join} reports the precision and the 
NDCG for $k$ from 10 to 50. \josie is omitted as it returns exact answers. 
For most competitors, the general trend is that both precision and NDCG 
increase with $k$. \deepjoin always outperforms alternatives 
and exhibits outstanding generalizability (trained on 30k columns and tested 
on 1M columns). The best performance, with an average precision of 72\% and 
NDCG of 81\%, is observed when MPNet is equipped. \deepjoinmpnet is better 
than \deepjoindbert because MPNet is pre-trained on a larger corpora and 
under a unified view of masked language modeling and permuted language 
modeling. 
% and its advantage over non-\deepjoin methods is consistent: the gap to the 
% runner-up is in the range of [0.036, 0.056] and [0.031, 0.053] for precision 
% and NDCG on Webtable and [0.036, 0.057] and [0.020, 0.031] for precision and 
% NDCG on Wikitable. 
\lshensemble's performance is mediocre due to the conversion from overlap 
condition to Jaccard condition, which becomes very loose when the sizes of 
query and target significantly differ. 
\revise{
For embedding methods, 
\turl is better than \tabert because the pre-trained tasks (column type annotation, etc.)
of \turl are closer to joinable table discovery than 
\tabert's question answering. 
Nonetheless, these tasks still significantly differ from joinable table 
discovery, and thus both are in general no better than \fasttext and \bert. 
Another reason why \tabert and \turl exhibit inferior performance is 
due to the limited data for pre-training; e.g., \turl is pre-trained on 
entity-focused Wikipedia tables.}
\fasttext is better than \bert and \mpnet, 
indicating that simply using PLMs without fine-tuning does not translate to 
higher accuracy than context-insensitive word embeddings. \mlp roughly 
performs the best among the methods other than \deepjoin, showing that a 
regression on top of word embeddings further improves the performance. 

For semantic join, Table~\ref{tab:exp:pex:unsupervised} reports the precision 
and NDCG evaluated under \pexeso's definition 
(Definition~\ref{def:semantic-joinability}). 
% \confversion{Since \deepjoinmpnet always 
% outperforms \deepjoindbert, in the rest of the experiments, we only show the 
% results equipped with MPNet for \deepjoin.} 
% For the performance of \deepjoin, \deepjoinmpnet still consistently 
% outperforms \deepjoindbert. 
\deepjoinmpnet reports an average precision of 91\% and NDCG of 75\%, delivering 
higher accuracy than alternatives for all the settings. 
\fasttext is competitive on Webtable but is not good on Wikitable. 
% On Webtable, 
% though the gap to the runner-up method is small for precision ([0.006, 0.032]), 
% it is more substantial in terms of NDCG ([0.033, 0.062]). On Wikitable, the 
% advantage of \deepjoin is even more significant, with a gap to the runner-up of 
% [0.215, 0.270] for precision and [0.165, 0.181] for NCDG. 
% For the alternative methods, \fasttext 
% is better on Webtable and \lshensemble is better on Wikitable. 
We also change the threshold $\tau$ for vector matching to 0.8 and 0.7, 
and report the accuracy in Tables~\ref{tab:exp:pex:unsupervised:tau0.8} and 
\ref{tab:exp:pex:unsupervised:tau0.7}, respectively. \deepjoinmpnet is still the 
best for low $\tau$ settings, though its precision and NDCG generally drop with 
$\tau$. Such trend is also observed in most other methods. This is because a 
lower $\tau$ suggests that more cell values are regarded as matching, and thus 
it tends to introduce less similar contents to the training examples, which are 
harder to deal with. 

We then evaluate these methods using the labels from our database researchers. The 
precision, recall, and F1 score when $k = 10$ are reported in Table~\ref{tab:exp:pex:human}. 
\deepjoinmpnet still performs the best. It is even better than \pexeso, and the 
advantage is remarkable, by a margin of 0.105 -- 0.165 in F1 score. We believe there 
are two reasons. First, \deepjoinmpnet uses a fine-tuned PLM, which captures the 
semantics of table contents in a better way than \pexeso which uses fastText to 
embed cell values. Second, \pexeso defines matching cells with a threshold. 
When judged by experts for joinability, the matching condition may differ across 
cell values, queries, and target columns, whereas a fixed threshold may not fit 
all of them. 
\revise{
For a detailed comparison, we show three typical win/lose examples in 
Table~\ref{tab:win-lose-example}. ``Win'' means only \deepjoin is able to correctly 
identify this entry, while ``lose'' means a false positive for \deepjoin but a true 
negative for at least one other competitor. The first example, which pertains to 
headboards, shows that the attention mechanism focuses on the word ``headboard'' in 
the table title and the words indicating bed size in the cell values. The second 
example, which pertains to tall buildings, shows that the PLM captures the similarity 
between the titles of the query and the target, as well as the building names in 
their contents. The third example, a false positive of \deepjoin, is potentially due 
to incorrect alignment of metadata. While these examples suggest that PLMs perform 
better than context-insensitive word embeddings when there are phrases indicating 
strong joinability, we also observe opportunities for improvement. 
}

\revise{
To drill down the cases of semantic joins, we randomly sample 100 query columns from 
the original corpus and request our experts to label the search results ($k = 10$) of 
four methods, \lshensemble, \fasttext, \pexeso, and \deepjoinmpnet. We divide the 
results into two cases: joins for data cleaning (near duplicates) references and joins 
for related columns (attribute enrichment). In Webtable, there are 68 tables for near 
duplicates and 177 tables for attribute enrichment. In Wikitable, there are 164 tables 
for near duplicates and 330 tables for attribute enrichment. We report the recalls of 
the four competitors in Table~\ref{tab:join-case-drill-down}. 
In general, lower recalls show that attribute enrichment is harder than near duplicates. This 
is expected, because for attribute enrichment, the matching condition is looser, meaning that 
we need to consider more columns that can be joined in a semantic manner. Nonetheless, 
\deepjoinmpnet exhibits superior performance in both cases, and the gap to the runner-up 
competitors are remarkable, especially for near duplicates, wherein the recall is around 
twice as much as the runner-up's. 
}

To investigate how the performance changes with column size, we divide target columns 
of Webtable into three groups according to their size: short (5 -- 10 cells), medium 
(11 -- 50 cells), and long ($>$ 50 cells). We only perform this experiment on Webtable 
because the number of columns in the long group is too small on Wikitable. For each 
group, we ensure that the query length is in the same range, and report the results in 
Table~\ref{tab:exp:accuracy-column-size}. For all the methods, the accuracy decreases 
with column size. This is because each column is transformed to a fixed-length object 
(MinHash sketch or vector). From the information perspective, the object after 
transformation has redundancy for short columns, but is compressed and more lossy for 
long columns. Nonetheless, \deepjoinmpnet is always the best, in line with what we 
have witnessed in the above experiments. 

\revise{We also perform an 
experiment on a synthetic dataset with 512 to 4,096 rows, in order to investigate the case 
when the input sequence length exceeds the \textsf{max\_seq\_length} limit of PLMs (e.g., 
512 tokens for \deepjoin). We synthesize 10k columns of 5 attributes: address, 
company, job, person, and profile, by using Faker~\cite{faker} with the default parameters 
that match real-world English word frequencies. We randomly take 50 columns as queries and 
the others are targets. The method that samples the most frequent cell values within 
\textsf{max\_seq\_length} tokens (see Section~\ref{sec:embed-column}), is dubbed 
-\textsf{frequency}. For comparison, we consider another two options: -\textsf{random}, 
which randomly samples cell values with no more than \textsf{max\_seq\_length} tokens, and 
-\textsf{truncate}, which truncates to the first \textsf{max\_seq\_length} tokens. The 
results are reported in Table~\ref{tab:exp:accuracy-tall-columns}. \deepjoinmpnet still 
consistently outperforms other models. For the three sampling options, -\textsf{random} is 
generally better than -\textsf{truncate}, and -\textsf{frequency} is always the best, 
justifying our argument that frequent cell values are more likely to yield join results.
}

\subsection{Ablation Study}
\label{sec:exp:ablation}
% \confversion{For ablation study, we report the results for equi-joins here and provide the 
% results for semantic joins in the extended version~\cite{extended-paper}.}

We first evaluate the impact of column-to-text transformation and test the seven 
options in Table~\ref{tab:column-to-text}. The results are reported 
%\confversion{in Table~\ref{tab:exp:col2text:equi}.}
in Tables~\ref{tab:exp:col2text:equi} and \ref{tab:exp:col2text:semantic}. 
Adding column name at the beginning 
(\texttt{colname-col}) improves the performance of simply concatenating cell 
values (\texttt{col}). Adding table title at the beginning 
(options with \texttt{title}) also has a positive impact. Appending statistical 
information (options with \texttt{stat}) further improves the performance, whereas 
appending context (options with \texttt{context}) has a negative impact. The 
latter is because the context includes information irrelevant to the column. Among 
the seven options, \texttt{title-colname-stat-col} is the best. 
% This is expected, as it contains the metadata of table and column, the content of 
% column, and statitical information. 

We then evaluate the impact of cell shuffle for data augmentation. We vary the 
shuffle rate (defined in Section~\ref{sec:training-data}) and report the results 
%\confversion{in Table~\ref{tab:exp:data-aug:equi}.}
in Tables~\ref{tab:exp:data-aug:equi} and \ref{tab:exp:data-aug:semantic}. \texttt{0.0} 
means there is no shuffle. We observe that a moderate shuffle rate achieves 
the best performance 
%\confversion{(0.2 for Webtable and 0.3 for Wikitable),} 
(0.2 and 0.3 for equi-joins and semantic joins on Webtable, 
respectively, and 0.3 and 0.4 for equi-joins and semantic joins on Wikitable, respectively), 
indicating that shuffling the cells in columns helps the model learn that the 
joinability is order-insensitive. On the other hand, over-shuffling is negative and 
even worse than no shuffle. We suspect this is because the original order of cells 
in both datasets follows some distribution. The attention mechanism in the PLM can 
capture such distribution and focus on the cells that are more probable to match. 
When the order is too random, the attention mechanism loses focus and thus a 
detrimental impact is observed. 

% \confversion{
% \begin{table}[t]
%   \centering
%   \caption{Processing time per query, varying $\size{\mathcal{X}}$, $k = 10$.}
%   \resizebox{\linewidth}{!}{%
%   \begin{tabular}{ l c | ccccc} 
%     \hline
%     & query encoding (ms) & \multicolumn{5}{c}{total (ms)} \\
%     Methods               &    & $\size{\mathcal{X}}$ = 1M &  2M &  3M &  4M & 5M   \\
%     \hline
%     \multicolumn{7}{c}{Webtable, equi-joins}\\ \hline
%     \lshensemble          & -  & 508 & 597 & 634 & 689 & 785    \\
%     \josie                & -  & 506 & 751 &  874 & 980 & 1103   \\
%     \fasttext             & 9 & 9.7 & 10.3 & 11.5 & 12.1 & 13.6  \\
%     \deepjoin (CPU)       & 66 & 68.1 & 69.3 & 71.4 & 73.2 & 74.1   \\
%     \deepjoin (GPU)       & \textbf{7} & \textbf{8.0} & \textbf{8.7} & \textbf{9.6} & \textbf{10.8} & \textbf{10.7} \\
%     \hline
    
%     \multicolumn{7}{c}{Webtable, semantic joins}\\ \hline
%     \pexeso               & -  & 2566 & 3116 & 3780 & 4122 & 4590   \\
%     \deepjoin (CPU)       & 74 & 76.1 & 77.9 & 78.4 & 80.1 & 79.9   \\
%     \deepjoin (GPU)       & \textbf{7}  & \textbf{8.4} & \textbf{8.8} & \textbf{9.5} & \textbf{9.7} & \textbf{10.9} \\
%     \hline
%     \end{tabular}
%   }
%   \label{tab:exp:time-dataset-size}
% \end{table}
% }

\begin{table}[t]
  \centering
  \caption{Processing time per query, varying $\size{\mathcal{X}}$, $k = 10$.}
  \resizebox{\linewidth}{!}{%
  \begin{tabular}{ l | c | ccccc} 
    \hline
    & query encoding (ms) & \multicolumn{5}{c}{total (ms)} \\ \hline
    Methods               &    & $\size{\mathcal{X}}$ = 1M &  2M &  3M &  4M & 5M   \\
    \hline
    \multicolumn{7}{c}{Webtable, equi-joins}\\ \hline
    \lshensemble          & -  & 508 & 597 & 634 & 689 & 785    \\
    \josie                & -  & 506 & 751 &  874 & 980 & 1103   \\
    \fasttext             & 9 & 9.7 & 10.3 & 11.5 & 12.1 & 13.6  \\
    \deepjoin (CPU)       & 66 & 68.1 & 69.3 & 71.4 & 73.2 & 74.1   \\
    \deepjoin (GPU)       & \textbf{7} & \textbf{8.0} & \textbf{8.7} & \textbf{9.6} & \textbf{10.8} & \textbf{10.7} \\
    \hline
    
    \multicolumn{7}{c}{Webtable, semantic joins}\\ \hline
    \pexeso               & -  & 2566 & 3116 & 3780 & 4122 & 4590   \\
    \deepjoin (CPU)       & 74 & 76.1 & 77.9 & 78.4 & 80.1 & 79.9   \\
    \deepjoin (GPU)       & \textbf{7}  & \textbf{8.4} & \textbf{8.8} & \textbf{9.5} & \textbf{9.7} & \textbf{10.9} \\
    \hline
    %& query encoding (ms) & \multicolumn{5}{c}{total (ms)} \\
                          &    & $\size{\mathcal{X}}$ = 200k &  400k &  600k & 800k & 1M   \\
    \hline
    \multicolumn{7}{c}{Wikitable, equi-joins}\\ \hline
    \lshensemble          & -  & 236 & 338 & 467 & 514 & 652    \\
    \josie                & -  & 304 & 377 & 455 & 556 & 647   \\
    \fasttext             & 6  & 6.7  & 6.6 & 6.8 & 6.9  & 7.4  \\
    \deepjoin (CPU)       & 76  & 76.4  & 76.7 & 76.9 &  77.0 & 77.1   \\
    \deepjoin (GPU)       & \textbf{5} & \textbf{5.4} & \textbf{5.8} & \textbf{5.8} & \textbf{6.1} & \textbf{6.3} \\
    \hline
    
    \multicolumn{7}{c}{Wikitable, semantic joins}\\ \hline
    \pexeso               & 1665   & 1874  & 1995 & 2310 & 2551 & 2789   \\
    \deepjoin (CPU)       & 86  & 86.5  & 86.9 & 87.1 & 87.4 & 87.7   \\
    \deepjoin (GPU)       & \textbf{9}  & \textbf{9.5} & \textbf{9.6} & \textbf{10.1} & \textbf{10.3} & \textbf{10.5} \\
    \hline
    \end{tabular}
  }
  \label{tab:exp:time-dataset-size}
\end{table}

\begin{table}[t]
  \centering
  \caption{Processing time per query, varying $k$.}
  \resizebox{\linewidth}{!}{%
  \begin{tabular}{ l | c | ccccc} 
    \hline
    & query encoding (ms) & \multicolumn{5}{c}{total (ms)} \\ \hline
    Methods     &    & $k$ = 10 &  20 &  30 &  40 & 50   \\
    \hline
    \multicolumn{7}{c}{Webtable, equi-joins}\\ \hline
    \lshensemble    & - & 496 & 506 & 590 & 595 & 508    \\
    \josie                 & - & 535 & 556 &  578 & 580 & 506   \\
    \fasttext                          & 9 & 10.3 & 10.5 & 10.2 & 10.8 & 11.1  \\
    \deepjoin (CPU)                    & 66 & 67.1 & 67.1 & 67.1 & 67.2 & 68.1   \\
    \deepjoin (GPU)                    & 7 & 8.4 & 8.1 & 8.2 & 8.1 & 8.0  \\
    \hline
    \multicolumn{7}{c}{Webtable, semantic joins}\\ \hline
    \pexeso                  & - & 2345 & 2444 & 2356 & 2754 & 2566   \\
    \deepjoin (CPU)          & 74 & 75.6 & 76.8 & 76.1 & 75.8 & 76.1   \\
    \deepjoin (GPU)          & 7 & 8.1 & 8.3 & 8.0 & 8.2 & 8.4 \\
    \hline    
    \multicolumn{7}{c}{Wikitable, equi-joins}\\ \hline
    \lshensemble    & - & 652 & 720 & 715 & 678 & 736    \\
    \josie                 & - & 647 & 667 &  708 & 697 & 788   \\
    \fasttext                          & 6 & 7.4 & 7.2 & 7.8 & 7.3 & 7.7  \\
    \deepjoin (CPU)                    & 76 & 77.1 & 78.1 & 77.4 & 77.5 & 77.6   \\
    \deepjoin (GPU)                    & 5 & 6.3 & 7.0 & 6.6 & 6.7 & 6.4  \\
    \hline
    \multicolumn{7}{c}{Wikitable, semantic joins}\\ \hline
    \pexeso                  & - & 2655 & 2776 & 2557 & 2743 & 2789   \\
    \deepjoin (CPU)          & 86 & 87.4 & 87.3 & 87.1 & 87.2 & 87.7  \\
    \deepjoin (GPU)          & 9 & 10.5 & 11 & 10.2 & 10.7 & 10.4 \\
    \hline
    \end{tabular}
  }
  \label{tab:exp:time-k}
\end{table}

\begin{table}[t]
  \centering
  \caption{Processing time per query, varying $\size{X}$, $k = 10$.}
  \resizebox{\linewidth}{!}{%
  \begin{tabular}{ l | ccc | ccc} 
    \hline
    & \multicolumn{3}{c|}{query encoding (ms)} & \multicolumn{3}{c}{total (ms)} \\ \hline
    Methods               & $\size{X}$ = 5 -- 10 &  11 -- 50 & $>$ 50  & $\size{X}$ = 5 -- 10 &  11 -- 50 & $>$ 50  \\
    \hline
    \multicolumn{7}{c}{Webtable, equi-joins}\\ \hline
    \lshensemble          & -  & -  & -  & 455  & 487  & 467  \\
    \josie                & -  & -  & -  & 410  & 589  & 792  \\
    \fasttext             & 5  & 6  & 6  & 5.8  & 6.7  & 6.9  \\
    \deepjoin (CPU)       & 71 & 75 & 78 & 71.7 & 75.4 & 78.5 \\
    \deepjoin (GPU)       & 4  & 5  & 6  & 4.9  & 5.8  & 6.9  \\
    \hline
    
    \multicolumn{7}{c}{Webtable, semantic joins}\\ \hline
    \pexeso               & -  & -  & -  & 2123 & 2785 & 3244 \\
    \deepjoin (CPU)       & 81 & 84 & 89 & 81.9 & 84.7 & 89.3 \\
    \deepjoin (GPU)       & 8  & 9  & 9  & 8.8  & 9.6  & 10.0 \\
    \hline
    \end{tabular}
  }
  \label{tab:exp:time-column-size}
\end{table}

\begin{table} [t]
  \small
  \centering
  \caption{\revise{Evaluation of embedding metrics, precision@10.}}
  %\resizebox{\linewidth}{!}{%
  \begin{tabular}{l|cc|cc} \hline
  & \multicolumn{2}{c|}{\deepjoindbert} & \multicolumn{2}{c}{\deepjoinmpnet} \\ \hline
    \multicolumn{5}{c}{Webtable, equi-joins} \\ \hline
    \backslashbox{train}{search} & \textsf{cosine} & \textsf{Euclidean} & \textsf{cosine} & \textsf{Euclidean} \\ \hline 
    %\textsf{dot}                     & 0.699 & 0.701 & 0.774 & 0.771 \\ 
    \textsf{cosine}                     & 0.701 & \textbf{0.702} & 0.730 & {0.731} \\
    \textsf{Euclidean}                  & 0.697 & 0.700 & 0.732 & \textbf{0.732} \\ \hline
    \multicolumn{5}{c}{Wikitable, equi-joins} \\ \hline
    \backslashbox{train}{search} & \textsf{cosine} & \textsf{Euclidean} & \textsf{cosine} & \textsf{Euclidean} \\ \hline 
    %\textsf{dot}                     & 0.585 & 0.586 & 0.621 & 0.620 \\ 
    \textsf{cosine}                     & \textbf{0.588} & \textbf{0.588} & \textbf{0.614} & \textbf{0.614} \\
    \textsf{Euclidean}                  & 0.587 & 0.585 & 0.611 & 0.612 \\ \hline
    \multicolumn{5}{c}{Webtable, semantic joins} \\ \hline
    %& \multicolumn{2}{c|}{\deepjoindbert} & \multicolumn{2}{c}{\deepjoinmpnet} \\ \hline
    \backslashbox{train}{search} & \textsf{cosine} & \textsf{Euclidean} & \textsf{cosine} & \textsf{Euclidean} \\ \hline 
    %\textsf{dot}                     & 0.857 & \textbf{0.862} & 0.871 & \textbf{0.874} \\ 
    \textsf{cosine}                     & \textbf{0.861} & \textbf{0.861} & 0.870 & \textbf{0.874} \\
    \textsf{Euclidean}                  & 0.859 & 0.859 & 0.870 & 0.866 \\ \hline
    \multicolumn{5}{c}{Wikitable, semantic joins} \\ \hline
    \backslashbox{train}{search} & \textsf{cosine} & \textsf{Euclidean} & \textsf{cosine} & \textsf{Euclidean} \\ \hline 
    %\textsf{dot}                     & \textbf{0.788} & 0.781 & 0.812 & \textbf{0.815} \\ 
    \textsf{cosine}                     & 0.787 & \textbf{0.788} & 0.810 & \textbf{0.813} \\
    \textsf{Euclidean}                  & 0.785 & 0.783 & 0.808 & 0.810 \\ \hline
  \end{tabular}
  %}
  \label{tab:train-search-metric}
\end{table}

\revise{
For the column embedding metrics used in offline training and online searching, 
we perform an evaluation of two metrics: cosine similarity and Euclidean distance. 
Since the Faiss library does not support cosine similarity for ANNS, we normalize 
column embeddings before ANNS and use the inner product as the metric, so as to output 
the same kNN results as using cosine similarity. 
The precisions when $k = 10$ are reported in Table~\ref{tab:train-search-metric}. 
Cosine similarity is better for training, while Euclidean distance is better for 
searching. The combination of cosine similarity for training and Euclidean distance 
for searching is overall the best. Since the performances of the two metrics are 
very close, users may address the discrepancy by choosing the same metric for 
training and searching. Nonetheless, we still use the best combination in our 
experiments. 
}

\subsection{Efficiency Evaluation}
\label{sec:exp:efficiency}
We vary the number of target columns and report the average query processing time  
%\confversion{on Webtable in Table~\ref{tab:exp:time-dataset-size}.} 
in Table~\ref{tab:exp:time-dataset-size}. 
For embedding methods, we also report query encoding time, which includes 
column-to-text transformation and column embedding. \josie and \pexeso are the 
slowest for equi-joins and semantic joins, respectively, and both exhibit 
substantial growth of search time (e.g., around 2 times when we increase the Webtable 
size from 1M to 5M). \lshensemble is also slow, despite transforming columns to 
fixed-length sketches. In contrast, embedding methods are much faster, though the 
majority of time is spent on query encoding. For example, \deepjoin (with MPNet), 
even if equipped with a CPU, is 7 -- 57 times 
(Webtable) and 3 -- 32 times (Wikitable) 
faster than the above methods. The growth of search time is also slight (e.g., 1.09 
times for equi-joins and 1.05 times for semantic joins, with Webtable's size from 1M 
to 5M), showcasing its scalability. With the help of a GPU, 
\deepjoin is substantially accelerated and can be 103 and 421 times faster than 
\josie and \pexeso, respectively, and even faster than \fasttext. 

Table~\ref{tab:exp:time-k} reports the query processing time when we vary $k$ from 
10 to 50. The general trend is that we spend more time for a larger $k$. 
Nonetheless, the growth for \deepjoin is very slight, because most of its overhead is 
query encoding, which is independent of the choice of $k$. As such, we roughly observe 
a greater speedup over existing methods when we increase $k$ from 10 to 50. 

To evaluate how the efficiency changes with column size, we use the same 
setting as in the corresponding 
% divide the target columns of Webtable into three groups and ensure the query lengths 
% are in the same range, in line with the corresponding 
accuracy evaluation 
(Table~\ref{tab:exp:accuracy-column-size}). Additionally, we sample and index only 300k  
target columns for each group, in order to eliminate the impact of the number of target 
columns. The results are reported in 
Table~\ref{tab:exp:time-column-size}. The exact methods, \josie and \pexeso, exhibit 
considerable growth (1.9 and 1.5 times, respectively) of query processing time when we 
switch from short to long columns, which reflects the analysis in 
Section~\ref{sec:state-of-the-art}. In contrast, the growth for embedding methods 
is much slighter. For example, we only observe a growth of 1.09 times for \deepjoin with 
a CPU, and this only affects its query encoding rather than the ANNS. For \deepjoin with 
a GPU, we also observe a more remarkable speedup over the exact methods on longer 
columns.

\section{Related Work}
\label{sec:related}

\mysubsection{Table discovery and data lake management}
Besides joinable table discovery~\cite{josie, lsh-ensemble}, techniques 
have been developed for searching unionable tables~\cite{lsh-table-union}. 
Another important problem is related table discovery. SilkMoth~\cite{silkmoth} 
models columns as sets and finds related sets under maximum bipartite 
matching metrics. JUNEAU~\cite{juneau} finds related tables for data science 
notebooks using a composite score of multiple similarities. D$^3$L~\cite{d3l} 
is a dataset discovery method which finds \topk results with a scoring function 
involving multiple attributes of a table. DLN~\cite{dln} discovers related 
datasets by exploiting historical queries having join clauses and determines 
relatedness with a random forest. Nextia$_{JD}$~\cite{nextiajd} uses 
meta-features (cardinalities, value distribution, entropy, etc.) to provide a 
join quality ranking of candidate columns. \revise{EMBER~\cite{ember} is a context 
enrichment system for ML pipelines, leveraging transformer-based~\cite{transformer}  
representation learning to automate keyless joins. 
% , which quantify the relatedness of records across datasets to identify records 
% of similar entities. 
Another notable work is Valentine~\cite{valentine}, in which an experiment suite 
was proposed for the experiments of dataset discovery with joinability and unionability.} 

% Set similarity have also been used to find related tables. For query 
% processing algorithms, we refer readers to \cite{set-similarity-survey} for 
% experimental comparison and \cite{pigeonring, sizeaware-overlap} for recent advances. 
% Xiao \etal proposed the pp-join algorithm~\cite{prefix-filter-tods11} 
% and its \topk version~\cite{prefix-filter-icde09}; Deng \etal proposed 
% a partition-based method \cite{partition-set-sim-join}. 
% Wang \etal designed a fuzzy join predicate that combines token and characters and 
% proposed the corresponding algorithm~\cite{fuzzy-token}. Wang \etal proposed 
% MF-join \cite{mf-join} that performs a fuzzy match with multi-level filtering. 
% The above solutions were not designed for data lakes (see \cite{josie}) 
% and only deal with raw textual data \revise{rather than high-dimensional vectors}. 

For data lake management, another problem which has been extensively studied 
is column type annotation. Notable approaches include Sherlock~\cite{sherlock}, 
Sato~\cite{sato}, and DODUO~\cite{doduo}. Among the three, DODUO is the one 
that employs PLMs. To deal with the case when tables differ in format, 
transformation techniques are often used to convert data so they can be joined. 
To tackle this problem, auto-join~\cite{auto-join} joins two tables with string 
transformations on columns. A similar method is 
auto-transform~\cite{auto-transform}, which learns string transformations with 
patterns. Besides, SEMA-join~\cite{sema-join} finds related pairs between two 
tables with statistical correlation. 
% Although optimizations for joining two given tables were introduced in these 
% studies, when applying to our problem, it is prohibitive to try joining every 
% table in the data lake with the query table.
\revise{Other representative problems include data lake 
organization~\cite{data-lake-organization}, data validation~\cite{data-validation}, 
and data lake integration~\cite{data-lake-integration}. A recent advancement 
is regarding data integration as prompting tasks for foundation 
models~\cite{foundation-model}.}

\mysubsection{Table embedding}
Language models have been used in understanding the contents of 
tables. For example, cell classification was investigated in 
\cite{cell-embedding}, with an RNN-based cell embedding method proposed. 
Table2Vec~\cite{table2vec}, featuring a series of embedding approaches 
for words, entities, and headers based on the idea of skip-gram model of 
word2vec~\cite{word2vec}, deals with the table retrieval problem that 
returns a ranked list of tables for a keyword query. Besides, PLMs such 
as BERT~\cite{bert} have also been used for table retrieval~\cite{bert4tr}. 
% accompanied by an embedding-based feature selection technique to select 
% relevant contents to a query and a multi-layer perceptron for relevance 
% score computation. 

More advanced approaches enlarged the scope of downstream tasks to include 
entity linkage, column type annotation, cell filling, etc., and designed 
pre-trained models that can be fine-tuned for them. TURL~\cite{turl} 
features a contextualization technique to convert table contents to sequences 
and leverages a masked language model (MLM) initialized by 
TinyBERT~\cite{tinybert}. TaPas~\cite{tapas} is built upon a similar MLM but 
employs BERT~\cite{bert}, with additional information embedded such as 
positions and ranks. TaBERT~\cite{tabert} pre-trains for question answering 
tasks and learns embeddings for cells, columns, and utterance tokens in the 
questions using BERT. By adopting two transformers to 
independently encode rows and columns, TABBIE~\cite{tabbie} embeds cells, 
columns, and rows, and achieves one order of magnitude less training time 
than TaBERT. %As a tree-based model using BERT, 
TUTA~\cite{tuta} creates trees 
to encode the information in hierarchical tables, while most previous studies 
focused on flat tables. Another method for hierarchical tables is 
GTR~\cite{gtr}, which models cells, rows, and columns as nodes in a graph and 
employs a graph transformer~\cite{graph-transformer} to capture the 
neighborhood information of cells, rows, and columns. 
%and the global information of rows and columns. 

% \mysubsection{Recommender systems}
% Recommender systems build a model from users past behaviors and suggest items 
% to users in a personalized manner. Early approaches are mainly based on 
% collaborative filtering~\cite{collaborative-filtering}, which suggests similar 
% items to users with similar past behaviors, or content-based 
% filtering~\cite{content-based-filtering}, which makes suggestions using the 
% pre-tagged characteristics of items. Factorization 
% machine~\cite{factorization-machine} exploits the interactions between features 
% and has become a popular model in the last decade. Recent approaches learn user 
% and item embeddings and adopt a multi-stage stragety that retrieve a candidate 
% set first and then rank/rerank the candidates. Such strategy has been widely 
% used in industrial solutions, such as YouTube~\cite{youtube-recsys}, 
% Facebook~\cite{facebook-recsys}, and Alibaba~\cite{alibaba-recsys}, and is often 
% accompanied by a two-tower model~\cite{two-tower}, which consists of two neural 
% networks for producing user and item embeddings, respectively. 
\section{Conclusion}
\label{sec:concl}
We proposed \deepjoin, a deep learning model that fits both equi- and 
semantic-joinable table discovery in a data lake. \deepjoin was designed 
in an embedding-based retrieval fashion, which embeds columns with a 
fine-tuned PLM and resorts to ANNS to find joinable results, thereby 
achieving a search time sublinear in the repository size. 
% A metric 
% learning approach was proposed so that columns are expected to be 
% joinable if they are close in the embedding space. 
% We devised techniques for contextualization and training data preparation. 
The experiments demonstrated the generalizability of \deepjoin, which is 
consistently more accurate than alternatives methods, as well as the superiority 
of \deepjoin in search speed, which is up to two orders of magnitude 
faster than alternatives. 
We expect that by employing more advanced 
retrieval strategies or PLMs, the performance of \deepjoin can be further 
improved. 

\begin{acks}
This work is mainly supported by NEC Corporation and partially 
supported by JSPS Kakenhi 22H03903 and CREST JPMJCR22M2.
\end{acks}

\clearpage

\balance
\bibliographystyle{abbrv}
\bibliography{main}

\end{document}